\documentclass[12pt,a4paper]{article}
\usepackage{jheppub}

\definecolor{klgreen}{rgb}{0.0, 0.5, 0.0}

\definecolor{color1}{rgb}{1,0,0}
\definecolor{color2}{rgb}{0,0,1}
\definecolor{color3}{rgb}{0,1,0}
\definecolor{color4}{rgb}{0.6,0.4,0.2}
\definecolor{color5}{rgb}{1,0,1}
\definecolor{color6}{rgb}{1,0.5,0}
\definecolor{color7}{rgb}{0.666667,0.666667,1}
\definecolor{color8}{rgb}{0.5,0,0.5}
\definecolor{color9}{rgb}{0.666667,0.666667,0}
\definecolor{green2}{rgb}{0.0, 0.5, 0.0}
\definecolor{gray2}{rgb}{0.5, 0.5, 0.5}

\newcommand{\exclude}[1]{}

\newcommand{\beq}{\begin{equation}}
\newcommand{\eeq}{\end{equation}}
\newcommand{\bea}{\begin{eqnarray}}
\newcommand{\eea}{\end{eqnarray}}
\long\def\/*#1*/{}

\newcommand{\junk}[1]{}

\title{\Large Non-Hermitian Holographic Flows to Little Rip Cosmologies}
\author[1,2]{Daniel Are\'an,}
\author[1,2]{David Garcia-Fariña}
\author[1]{Juan F. Pedraza}
\affiliation[1]{Instituto de F\'isica Te\'orica UAM/CSIC, 28049 Madrid, Spain}
\affiliation[2]{Departamento de F\'isica Te\'orica, Universidad Aut{\'o}noma de Madrid, 28049 Madrid, Spain}

\preprint{\texttt{IFT-UAM/CSIC-26-86}}

\emailAdd{daniel.arean@uam.es}
\emailAdd{david.garciafarinna@estudiante.uam.es}
\emailAdd{j.pedraza@csic.es}

\usepackage{physics} 
\usepackage{tensor} 
\usepackage{subcaption}
\usepackage{graphicx}

\abstract{
Spacelike singularities supported by matter satisfying the null energy condition (NEC) are expected to fall within the Belinski--Khalatnikov--Lifshitz (BKL) paradigm. We show that controlled violations of the NEC in holography can lead to different black hole interiors. In a holographic model dual to a strongly coupled non-Hermitian $\mathcal{PT}$-symmetric QFT, we uncover a novel non-Kasner regime within its real, $\mathcal{PT}$-restored phase. The deep-interior geometry describes an isotropic FLRW cosmology undergoing super-accelerated expansion and approaching a Little Rip. This regime leaves a characteristic imprint on two-sided heavy-operator correlators, allowing it to be distinguished from a standard Kasner interior. Our construction provides a concrete holographic realization of a Little Rip cosmology and lays the groundwork for a Little Rip/CFT correspondence, in which such cosmological regimes can be explored through observables in a non-Hermitian quantum field theory.
}

\begin{document}

\maketitle

\section{Introduction}

Cosmological singularities are a ubiquitous feature of solutions to the Einstein equations, and considerable effort has therefore been devoted to understanding and classifying their possible behavior. The Belinski-Khalatnikov-Lifshitz (BKL) paradigm \cite{Misner1969,Belinsky:1970ew,lifshitz1971asymptotic} predicts that, in the vicinity of a generic spacelike singularity, the gravitational dynamics becomes ultralocal: time derivatives in the Einstein equations dominate over spatial derivatives, and the geometry at each spatial point is approximately described by a Kasner metric. In comoving coordinates, this metric takes the form
\begin{equation}
ds^2=-d\tau^2+\sum_{i=1}^{d}\tau^{2p_i(x)}(dx^i)^2,
\end{equation}
where the functions $p_i(x)$ are the local Kasner exponents.\footnote{While ultralocality was originally proposed as a central principle of BKL dynamics, the Carroll expansion of GR has recently provided a covariant, off-shell realization of this limit, together with a systematic framework for understanding the emergence of chaotic near-singularity dynamics~\cite{Oling:2024vmq}.} This Kasner form is not expected to remain valid throughout the entire approach to the singularity. Rather, in the oscillatory regime, the evolution consists of a sequence of Kasner epochs connected by rapid transitions, which can be described as billiard motion in an auxiliary hyperbolic space and give rise to the characteristic chaotic behavior of the BKL regime \cite{Damour:2000wm,Damour:2000hv,Damour:2002et,Henneaux:2007ej}. Remarkably, the Kasner/BKL paradigm is robust for broad classes of conventional matter sectors satisfying the null energy condition (NEC): the approach to a generic spacelike singularity remains locally Kasner-like, possibly connected by rapid transitions between successive Kasner epochs \cite{Andersson:2000cv,Damour:2002tc}.

In the context of AdS/CFT, BKL singularities have garnered considerable attention due to their potential imprints on the dual field theory. In particular, the near-singularity region is not entirely screened from the boundary description: although it lies behind the event horizon, its geometry can control distinctive non-analytic structures and asymptotic regimes of CFT observables. Semiclassical two-sided correlators of heavy operators provide a prominent example, as their analytically continued behavior can encode properties of the deep interior \cite{Fidkowski:2003nf,Festuccia:2005pi}. Complementary candidate probes, including suitably chosen complexity=anything functionals \cite{Jorstad:2023kmq,Arean:2024pzo} and the thermal $a$-function \cite{Caceres:2022smh}, can likewise retain sensitivity to the near-singularity scaling and thereby distinguish between different Kasner regimes. These developments have motivated the study of asymptotically AdS black holes with Kasner-like interiors \cite{Frenkel:2020ysx,Hartnoll:2020rwq,Hartnoll:2020fhc,Sword:2021pfm,Sword:2022oyg,Wang:2020nkd,Mansoori:2021wxf,Liu:2021hap,Das:2021vjf,An:2022lvo,Auzzi:2022bfd,Mirjalali:2022wrg,Hartnoll:2022snh,Hartnoll:2022rdv,Caceres:2022hei,Liu:2022rsy,Gao:2023zbd,Caceres:2023zhl,Blacker:2023ezy,Caceres:2023zft,Carballo:2024hem,Caceres:2024edr}. Most known constructions, however, approach a final Kasner epoch or undergo at most a finite sequence of transitions, rather than realizing the infinite chaotic succession characteristic of BKL dynamics. The first explicit realization of genuine mixmaster chaos inside an AdS black hole was obtained in four dimensions in \cite{DeClerck:2023fax} and was subsequently generalized to arbitrary spacetime dimensions in \cite{Caceres:2026mug}.

The holographic constructions discussed above are supported by conventional bulk matter satisfying the NEC. A natural setting in which to explore qualitatively different interior dynamics 
can therefore be achieved by considering holographic models where the NEC is violated in a well-motivated manner. In this paper we study one such model in the context of non-Hermitian holographic theories.
A holographic model for a non-Hermitian $\mathcal{PT}$-symmetric theory was introduced in \cite{Arean:2019pom} and subsequently explored in \cite{Morales-Tejera:2022hyq,Xian:2023zgu,Arean:2024gks,Arean:2024lzz}.
Related holographic RG flows were studied in \cite{Begines:2026mlv} and shown to be connected to traversable wormhole geometries, while holographic duals of $\mathcal{PT}$-symmetric boundary CFTs were constructed in \cite{Maeda:2026awj}.
Remarkably, the model of \cite{Arean:2019pom} admits black hole solutions in which the bulk NEC is violated as a consequence of the non-Hermitian deformation on the dual field theory. This motivates us to investigate their interiors, since NEC violation opens the possibility of qualitatively new behavior beyond the standard BKL paradigm.

Indeed, we uncover a new non-Kasner asymptotic regime of the Little Rip type \cite{Frampton:2011sp}. The deep-interior geometry describes an isotropic cosmology undergoing accelerated expansion, thereby realizing a holographic flow from AdS to a Little Rip cosmology. Little Rip cosmologies were introduced in \cite{Frampton:2011sp}, 
and further developed in \cite{Brevik:2011mm,Frampton:2011rh,Brevik:2012nt},
as models of eternal accelerated expansion in which no singularity is reached at finite proper time. Instead, the curvature invariants diverge only asymptotically, as the proper time tends to infinity. Strictly speaking, the endpoint is therefore not a spacetime singularity in the sense of geodesic incompleteness, but rather an asymptotic curvature blow-up. For simplicity, however, we will henceforth refer to this regime as a singularity.

\begin{figure}
\centering
\includegraphics[width=0.49\textwidth]{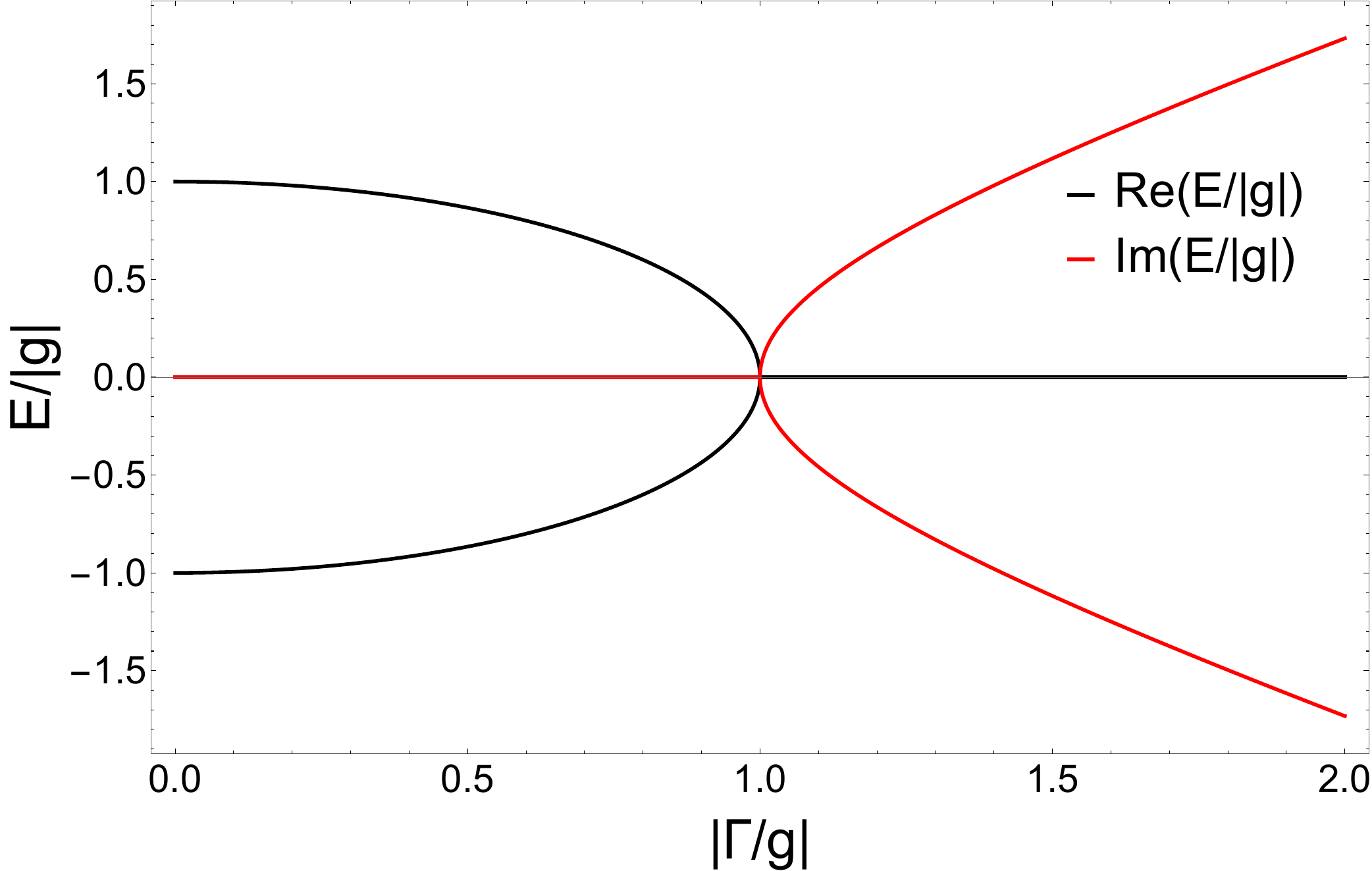}
\caption{Energy levels in the two-level system
\eqref{eq:H2level}. For $|\Gamma/g|>1$ the energies become complex conjugate and $\mathcal{PT}$ symmetry is broken.}
\label{fig:toymodel_2lvlEnergies}
\end{figure}

The interest in non-Hermitian $\mathcal{PT}$-symmetric quantum theories stems from the observation that Hermiticity of the Hamiltonian is not strictly necessary for a real energy spectrum and unitary time evolution. Instead, these properties can be recovered when the Hamiltonian possesses an unbroken antilinear symmetry, such as $\mathcal{PT}$ symmetry, together with an appropriate choice of inner product \cite{Bender:1998ke,Bender:2005tb,Mannheim:2015hto,Fring:2022tll}. In an open-system interpretation, non-Hermitian terms describe the exchange of matter or probability with an external environment. In the unbroken $\mathcal{PT}$ phase, gain and loss are balanced, allowing the system to reach a stationary state and yielding a real energy spectrum. By contrast, when $\mathcal{PT}$ symmetry is spontaneously broken, this balance is lost, the eigenvalues become complex, and the dynamics exhibits exponential growth or decay rather than stationary unitary evolution. These features are illustrated by the $\mathcal{PT}$-symmetric two-level Hamiltonian \cite{Bender:2005tb}
\begin{equation}
    H=\begin{pmatrix}-i \Gamma & g \\ g & i\Gamma\end{pmatrix}\,,
    \label{eq:H2level}
\end{equation}
reviewed in~\cite{Arean:2024gks}. This system features both a $\mathcal{PT}$-symmetric and a $\mathcal{PT}$-broken regime. $\mathcal{PT}$ symmetry is intact when the hopping between the
two levels, set by $g$, is faster than the inflow (outflow) of matter from (to) the exterior, fixed by $\Gamma$.
On the other hand, in the regime where the inflow (outflow) is faster than the hopping, no stationary state can be reached and $\mathcal{PT}$ is spontaneously broken. 
As shown in Fig.~\ref{fig:toymodel_2lvlEnergies}, the energy levels of the system are real in the $\mathcal{PT}$-symmetric regime and complex in the $\mathcal{PT}$-broken phase. 
They meet at the so-called exceptional point $|\Gamma/g|=1$.

The above features are realized in the holographic model of \cite{Arean:2019pom}, where a Hermitian CFT$_3$ is deformed by turning on constant non-Hermitian sources\footnote{The effects of spacetime-dependent sources were studied in \cite{Morales-Tejera:2022hyq,Arean:2024gks}.} for a charged scalar operator $\mathcal{O}$,
\begin{equation}
    S=S_\text{CFT}+\int d^3x\, \left( \bar{s} \mathcal{O}+s \bar{\mathcal{O}}\right)\,.
\end{equation}
From the bulk perspective, the non-Hermiticity enters only through the boundary conditions of the scalar field dual to $\mathcal{O}$. 
At zero temperature, when the Hermitian part of the source $s_h=s+\bar{s}$ is greater in absolute value than the non-Hermitian source $s_{nh}=s-\bar{s}$ the theory realizes a $\mathcal{PT}$-symmetric phase with real metric that satisfies the NEC. 
On the other hand, when the non-Hermitian source is larger 
the geometry becomes complex and $\mathcal{PT}$ is broken. The point $|s_h|=|s_{nh}|$ is an exceptional point. 

In practice, in the $\mathcal{PT}$-symmetric regime a non-Hermitian theory can be mapped to a Hermitian one via a Dyson map \cite{Mostafazadeh:2001jk,Mostafazadeh:2001nr,Mostafazadeh:2002id}. Nonetheless, in many cases, the mapping can be highly non-trivial and give rise to non-local interactions \cite{Bender:2004sa}.\footnote{The construction of the Dyson map for a QFT in a perturbative regime may require resummation of series in the coupling constant and be therefore highly non-trivial~\cite{Alexandre:2022uns}. Also, it is worth noting that, even if the isospectral Hermitian theory is local, one expects the Dyson map to be non-local in the presence of interactions \cite{Li:2024xms}.}
This alone motivates the study of non-Hermitian theories for their own physical relevance. Furthermore, once we rid ourselves of the constraints of Hermiticity and instead construct $\mathcal{PT}$-symmetric Hamiltonians, we can begin to study the spontaneous breaking of $\mathcal{PT}$ as a function of the parameters of the model. This is of particular interest since in the transition between $\mathcal{PT}$-symmetric and $\mathcal{PT}$-broken regimes there is an exceptional point where the Hamiltonian becomes non-diagonalizable.
Exceptional points exhibit a rich branch-point topology and give rise to a wide range of remarkable phenomena, including enhanced spectral sensitivity to perturbations~\cite{Wiersig:2014zbi,Chen:2017bjr} and unconventional transport behavior in photonic systems, such as nonreciprocal wave propagation and invisibility~\cite{PhysRevLett.106.213901,Feng:2012fjq,Peng:2013hoe} (see \cite{Ozdemir:2019iqe,Ashida:2020dkc,Ding:2022juv} for reviews). In the context of two-dimensional CFTs
exceptional points are expected to be described by logarithmic CFTs where the non-diagonalizable Hamiltonian is reflected in the appearance of Jordan blocks for the dilatation operator \cite{Gurarie:1993xq,Cardy:2013rqg,Grumiller:2010rm,Grumiller:2013at,Io:2026nxe}.

Since the pioneering paper \cite{Bender_2004}, non-Hermitian quantum field theories have attracted considerable interest.
It was shown that they may possess entirely real spectra in the unbroken $\mathcal{PT}$-symmetric regime~\cite{Bender:2004sa}.
More recently, non-Hermitian quantum field theories with spontaneous symmetry breaking, gauge symmetries, and supersymmetry have been constructed and studied in detail~\cite{Alexandre:2018uol,Mannheim:2018dur,Fring:2019hue,Alexandre:2019jdb,Fring:2019xgw,Alexandre:2020wki}. Vortices in non-Hermitian superfluids have also been recently considered in \cite{Begun:2021wol,Battye:2025seo}.

To connect these features of non-Hermitian field theories with the physics of black holes and their interiors, we must first specify the finite-temperature state. In standard AdS/CFT, an eternal two-sided black hole is dual to two copies of the boundary theory prepared in the thermofield double (TFD) state \cite{Maldacena:2001kr}. Throughout this work, we define its non-Hermitian analogue by taking both copies to be governed by the same non-Hermitian Hamiltonian, but with opposite orientations of time. From the bulk perspective, this choice is natural: it ensures that the two boundaries of the eternal black hole describe the same theory, up to time reversal. Importantly, the structure of this TFD state effectively restores $\mathcal{PT}$ symmetry even when the individual copies lie in the $\mathcal{PT}$-broken regime. By this we mean that observables evaluated in the unperturbed TFD state display no signatures of $\mathcal{PT}$ breaking. Nevertheless, the latter can manifest itself through instabilities of the TFD under perturbations. Physically, this suggests that finite-temperature effects can produce a finely balanced stationary state in which the gain and loss associated with the open-system description compensate each other. This balance is fragile, however, and small perturbations can drive the system away from equilibrium.

The above picture precisely mirrors the behavior found in the holographic model of \cite{Arean:2019pom}. At nonzero temperature, the model admits real solutions with $|s_{nh}|>|s_h|$ that are effectively $\mathcal{PT}$-symmetric but violate the bulk NEC and are dynamically unstable.\footnote{In \cite{Arean:2019pom}, the NEC-violating solutions were found to possess an unstable quasinormal mode in the charge-diffusion sector. Since the gauge field plays no role in constructing the background, this instability may be avoided by considering a theory in which the scalars are not coupled to the gauge field. Stable NEC-violating solutions have also been constructed in the non-Hermitian holographic models of \cite{Begines:2026mlv}.} Such solutions do not exist for arbitrarily large values of $\eta=|s_{nh}/s_h|$: there is a critical value $\eta_c$ beyond which real solutions cease to exist. For $\eta>\eta_c$, only complex solutions with complex temperatures were found in \cite{Xian:2023zgu}. By contrast, for $\eta<1$, the solutions are stable, preserve $\mathcal{PT}$ symmetry, and satisfy the NEC. Moreover, it was shown in \cite{Arean:2019pom} that a complexified $U(1)$ transformation maps these solutions to those of a Hermitian theory.

In this paper we will focus mainly on studying the black hole interiors of the real solutions with $|s_{nh}|>|s_{h}|$; which we will call $\mathcal{PT}$-restored solutions. As stated at the beginning of this section, these solutions are of particular interest as they violate the NEC while also being physically well
motivated from the perspective of an open system. This allows us to bypass the standard BKL paradigm and potentially find new behaviors in the vicinity of the singularity. Among the $\mathcal{PT}$-restored solutions there are two competing branches. The thermodynamically subdominant one is smoothly connected to the solutions with $\eta<1$ while the dominant one is not.

We find that the subdominant branch presents two qualitatively distinct behaviors. In one regime, the singularity is of Kasner type, with Kasner exponents continuously connected to those of the solutions with ($|\eta|<1$). In the complementary regime, as well as throughout the dominant branch, the black hole interior has a much richer structure, leading to a new type of singularity that does not belong to the Kasner class. Instead, we find an isotropic singularity described by an FLRW universe whose energy density grows as a power of the logarithm of the scale factor.
This geometry leaves a non-trivial imprint in heavy-operator geodesics. We find that the renormalized geodesic length as a function of energy $\mathcal{L}_{\text{ren}}(E)$
behaves non-algebraically in the large-$E$ limit and vanishes slower than any power law.
Recalling that in Kasner $\mathcal{L}_{\text{ren}}(E)$ follows a power law \cite{Frenkel:2020ysx}, this signature allows us to distinguish this new type of black hole interior from the standard BKL one. 

The paper is structured as follows. In section \ref{sec:TFD toy model} we discuss briefly in the context of a Quantum Mechanics toy model why $\mathcal{PT}$ symmetry is effectively restored for our choice of TFD. In section \ref{sec:Holographic model} we review the construction of the non-Hermitian holographic model of \cite{Arean:2019pom} and the corresponding phase diagram \cite{Xian:2023zgu}. In section \ref{sec:BHInteriors} we discuss the different black hole interiors we find for the different phases and in section \ref{sec:NewSingularity} we describe in detail the novel singularity found in the model. Lastly in section \ref{sec:Conclusions} we present our conclusions and comment on possible future directions. In the appendices we collect the holographic dictionary, discuss the numerical methods employed and present the analysis of the singularity for scalar potentials consisting of just a mass term. 

\section{TFD in a non-Hermitian toy model}\label{sec:TFD toy model}

To acquire some intuition about TFD states in non-Hermitian $\mathcal{PT}$-symmetric models we consider the $\mathcal{PT}$-symmetric two-level non-Hermitian Hamiltonian presented in the introduction
\begin{equation}\label{eq:PTHamiltonian_2LevelSystem}
    H=\begin{pmatrix}-i \Gamma & g \\ g & i\Gamma\end{pmatrix}\,,
\end{equation}
where $\Gamma$ and $g$ are real parameters. In this theory, $\mathcal{PT}$ is given by
\begin{equation}
    \mathcal{PT}=\begin{pmatrix}0  & 1 \\ 1 & 0\end{pmatrix}\mathcal{C}\,,
\end{equation}
with $\mathcal{C}$ the complex conjugation operator. 

When $|\Gamma/g|<1$, the eigenvalues of $H$ are real and the corresponding eigenstates are also eigenstates of $\mathcal{PT}$.
On the other hand, when $|\Gamma/g|>1$, the eigenvalues of $H$ are complex and the associated eigenstates are no longer eigenstates of $\mathcal{PT}$. Nonetheless, as $H$ commutes with $\mathcal{PT}$, the complex energies are related by complex conjugation and the eigenstates transform into each other under $\mathcal{PT}$.

Importantly, as $H$ is non-Hermitian, the associated eigenstates are not orthogonal, even in the $\mathcal{PT}$-symmetric regime. This feature implies that one should replace the notion of orthogonality by that of biorthogonality \cite{Brody:2013axr} such that matrix elements of an operator $A$ are given by $\bra{\tilde{n}}A\ket{m}$ where $\{\bra{\tilde{n}}\}$ is the dual basis, defined such that $\bra{\tilde{n}}\ket{m}=\delta_{nm}$. Importantly this implies that the trace of $A$ is defined as
\begin{equation}\label{eq:TraceDef}
    \Tr{A}=\sum_n  \bra{\tilde{n}}A\ket{n}\,,
\end{equation}
If $\{\ket{n}\}$ is the basis of right-eigenstates of the Hamiltonian, the dual basis $\{\bra{\tilde{n}}\}$ are the left-eigenstates 
\begin{equation}
    H\ket{n}=E_n\ket{n}\,,\qquad \bra{\tilde{n}}H=E_n\bra{\tilde{n}}\,,
\end{equation}
We note that, this formulation is equivalent to the standard one of pseudo-Hermitian Quantum Mechanics \cite{Mostafazadeh:2001jk,Mostafazadeh:2001nr,Mostafazadeh:2002id} by noting that in such case $\bra{\tilde{n}}=\bra{n}\gamma$ where $\gamma$ is the metric operator \cite{Mostafazadeh:2008pw,Fring:2022tll}.

With this setup in mind, let us consider a thermal density matrix 
\begin{equation}\label{eq:thermal dsty mtrx}
\rho_{th}=\frac{e^{-\beta H}}{\Tr{e^{-\beta H}}}=\frac{e^{-\beta H}}{\sum_n e^{-\beta E_n}}=\frac{e^{-\beta H}}{\mathcal{Z}}\,,
\end{equation}
with $\beta$ real and some generic operator $\mathcal{O}$ transforming non-trivially under $\mathcal{PT}$ to $\mathcal{O}^{\mathcal{PT}}=\mathcal{PT}\,\mathcal{O}\,\mathcal{PT}$. Given the non-trivial transformation properties of the operator, we can use it to diagnose the breaking of $\mathcal{PT}$; if $\mathcal{PT}$ is spontaneously broken we expect that for some choice of $\mathcal{O}$ we would have that 
\begin{equation}
    \Tr{\mathcal{O}\rho_{th}}\neq \Tr{\mathcal{O}^{\mathcal{PT}}\rho_{th}}^*\,,
\end{equation}
where the complex conjugation appears due to the anti-unitary nature of $\mathcal{PT}$. Remarkably,  what we see instead for the thermal state is that there are no signatures of $\mathcal{PT}$-breaking. 
Evaluating the right hand side of the equation for 
generic $\mathcal{O}$ in the regime where $|\Gamma/g|>1$ we have
\begin{align}\label{eq: Thermal expectation value OPT toy model}
    \Tr{\mathcal{O}^{\mathcal{PT}}\rho_{th}}^*=& \left\{\bra{\tilde{\psi}_+}\ket{\mathcal{O}^{\mathcal{PT}}\rho_{th}\,\psi_+}+\bra{\tilde{\psi}_-}\ket{\mathcal{O}^{\mathcal{PT}}\rho_{th}\,\psi_-}\right\}^*\nonumber\\
    =&\left\{ \frac{e^{-\beta E_+}}{\mathcal{Z}}\bra{\mathcal{PT}\tilde{\psi}_+}\ket{\mathcal{O}\,\mathcal{PT}\psi_+}^*+  \frac{e^{-\beta E_-}}{\mathcal{Z}}\bra{\mathcal{PT}\tilde{\psi}_-}\ket{\mathcal{O}\,\mathcal{PT}\psi_-}^*\right\}^*\nonumber\\
     =&\frac{e^{-\beta E_-}}{\mathcal{Z}}\bra{\tilde{\psi}_-}\ket{\mathcal{O}\,\psi_-}+  \frac{e^{-\beta E_+}}{\mathcal{Z}}\bra{\tilde{\psi}_+}\ket{\mathcal{O}\,\psi_+}\,=\Tr{\mathcal{O}\rho_{th}}\,,
\end{align}
where $H\ket{\psi_\pm}=E_\pm\ket{\psi_\pm}$, $E_+=E_-^*$ and we have used that $\mathcal{Z}$ is real and that $\mathcal{PT}\psi_\pm=\psi_\mp$. 
Hence, even when $\mathcal{PT}$ is broken for the eigenstates of $H$, the symmetry is restored for thermal states. In other words, turning on temperature allows us to reach a fine-tuned stationary state. 
This property can be shown to hold generally for density matrices of the form \eqref{eq:thermal dsty mtrx} with $\beta\in\mathbb{R}$. Indeed this feature was observed also in the holographic model of \cite{Arean:2019pom}.

Note that a generic density matrix does present signatures of $\mathcal{PT}$-breaking for $|\Gamma/g|>1$. The lack of signatures for the thermal density matrix is a consequence of its particular form. Nonetheless, even though the breaking of $\mathcal{PT}$ does not imprint itself in expectation values, it manifests itself in the instability of the thermal density matrix. Considering the evolution equation for a generic density matrix $\rho=\rho_{th}+\delta\rho e^{-i\omega t}$ 
\begin{equation}
    \frac{d}{dt}\rho=-i\left[H,\rho \right]\Rightarrow \omega\delta\rho=\left[H,\delta\rho\right]
\end{equation}
and rewriting it as an eigenvalue problem for $\omega$, we find that $\omega=\pm\sqrt{g^2-\Gamma^2}$. Hence when $|\Gamma/g|>1$ we have an unstable mode. Thus, the thermal state is unstable in the regime where the eigenstates of $H$ break $\mathcal{PT}$. Physically this implies that even a small deformation breaks the fine-tuned stationary state that appeared as a consequence of turning on temperature. 
Again this feature matches the observations of the holographic model of \cite{Arean:2019pom}.

Lastly, with 
the goal
of studying black hole interiors in the holographic setup, let us construct the TFD state for this toy model and comment on some salient features.
When constructing the TFD there is arbitrariness in how to double the system. 
We choose to double the system by considering an identical copy for which time evolves in the opposite direction. Hence, labeling the copies as $R$ (right) and $L$ (left),  we have the following TFD state 
\begin{equation}
    \ket{\text{TFD}}=\sum_n{\frac{e^{-\beta E_n/2}}{\sqrt{\mathcal{Z}}} \ket{n^R,n^L}}={\frac{e^{-\beta E_+/2}}{\sqrt{\mathcal{Z}}}  \ket{\psi_+^R,\psi_+^L}}+{\frac{e^{-\beta E_-/2}}{\sqrt{\mathcal{Z}}} \ket{\psi_-^R,\psi_-^L}}\,,
\end{equation}
such that it is invariant under time evolution with $H_-=H_R\otimes1_L-1_R\otimes H_L$ while it transforms non-trivially under $H_+=H_R\otimes1_L+1_R\otimes H_L$. The dual to this state is given by
\begin{equation}
    \bra{\tilde{\text{TFD}}}=\sum_n{\frac{e^{-\beta E_n/2}}{\sqrt{\mathcal{Z}}} \bra{\tilde{n}^R,\tilde{n}^L}}={\frac{e^{-\beta E_+/2}}{\sqrt{\mathcal{Z}}}  \bra{\tilde{\psi}_+^R,\tilde{\psi}_+^L}}+{\frac{e^{-\beta E_-/2}}{\sqrt{\mathcal{Z}}} \bra{\tilde{\psi}_-^R,\tilde{\psi}_-^L}}\,.
\end{equation}
Note that in holography this choice corresponds to taking a perspective where the second boundary describes the same non-Hermitian theory but with the direction of time reversed.

In the doubled system, $\mathcal{PT}$ is constructed by acting simultaneously with $\mathcal{PT}$ on the $R$ and $L$ subsystems, i.e. $\mathcal{PT}=\mathcal{PT}_R\otimes \mathcal{PT}_L$. Hence, given our construction, it follows that the TFD is invariant under $\mathcal{PT}$. As the TFD is the purification of the thermal state, we can now interpret the lack of signatures of $\mathcal{PT}$-breaking in the thermal state as a consequence of the $\mathcal{PT}$ symmetry of the TFD.

Despite the apparent $\mathcal{PT}$ symmetry of the TFD, there are still signatures of non-unitary behavior for $|\Gamma/g|>1$. In particular, expectation values of two-sided operators evolved with $H_+$ show oscillatory behavior in time for $|\Gamma/g|<1$ and exponential behavior in time for $|\Gamma/g|>1$.  Hence, by studying expectation values in the TFD state of two-sided operators evolved with $H_+$ we can predict when the eigenstates of the Hamiltonian cease to be invariant under $\mathcal{PT}$.

\section{Holographic model}\label{sec:Holographic model}

Following \cite{Arean:2019pom,Arean:2024lzz}, we construct a holographic non-Hermitian QFT by deforming a 3-dimensional Hermitian CFT with a non-Hermitian source of a relevant scalar operator charged under a global $U(1)$
\begin{equation}\label{eq:QFT NH action}
    \mathcal{S}_{QFT}=\mathcal{S}_{CFT}+\int d^3x \left[M \left(\mathcal{O}+\bar{\mathcal{O}}\right)+\eta\, M \left(\mathcal{O}-\bar{\mathcal{O}}\right) \right]\,,
\end{equation}
where $M,\eta\in\mathbb{R}$ and $\eta$ is a dimensionless constant encoding the strength of the non-Hermitian contribution. 
Proceeding as in~\cite{Arean:2024lzz},
defining the action of $\mathcal{PT}$ as
\begin{align}
    x=(t,x^1,x^2)\xrightarrow{\mathcal{PT}}\mathcal{PT}x=(-t,-x^1,x^2)\,,\\
    \mathcal{O}(x)\xrightarrow{\mathcal{PT}}\mathcal{O}(\mathcal{PT}x)\,,\qquad \bar{\mathcal{O}}(x)\xrightarrow{\mathcal{PT}}\bar{\mathcal{O}}(\mathcal{PT}x)\,,
\end{align}
the above theory is $\mathcal{PT}$-symmetric as long as $\mathcal{S}_{CFT}$ is $\mathcal{PT}$-symmetric.

For $|\eta|<1$, we can identify the Dyson map as a complexified global $U(1)$ transformation 
\begin{equation}\label{eq:DysonMapBdry}
\mathcal{O}\rightarrow e^{-i q \alpha} \mathcal{O}\,,\qquad \bar{\mathcal{O}}\rightarrow e^{i q \alpha} \bar{\mathcal{O}}\,,\qquad \alpha=\frac{i}{2q}\log\left(\frac{1-\eta}{1+\eta}\right) \,,
\end{equation}
where $q$ is the charge of $\mathcal{O}$.
Trivially under this similarity transformation the action \eqref{eq:QFT NH action} is mapped to the following Hermitian action
\begin{equation}
    \mathcal{S}_{QFT}=\mathcal{S}_{CFT}+\int d^3x \left[\tilde{M}\left(\mathcal{O}+\bar{\mathcal{O}}\right)\right]\,,
\end{equation}
with $\tilde{M}=M\sqrt{1-\eta^2}\in\mathbb{R}$ \cite{Arean:2019pom}. Hence, for $|\eta|<1$ the theory is trivially in the $\mathcal{PT}$-symmetric case. On the other hand, as shown in \cite{Arean:2019pom}, $\mathcal{PT}$ is spontaneously broken for $|\eta|>1$. Let us finally emphasize that we should not interpret the Dyson map as a symmetry of the theory, but rather as a duality between two distinct theories: one describing an open system and another a closed one. 

To capture these dynamics we follow \cite{Arean:2019pom} and consider the holographic model introduced in \cite{Gubser:2008wz}
characterized by the following Einsten-Maxwell-scalar action
\begin{equation}\label{eq:Bulk_Grav_Action}
    \mathcal{S}=\int d^4y \sqrt{-g}\left(R-2\Lambda-\frac{1}{4}F_{MN}F^{MN}-\mathcal{D}_M\phi \mathcal{D}^M\bar{\phi}-m^2\phi\bar{\phi}-v (\phi\bar{\phi})^2\right)\,,
\end{equation} 
where we work in dimensionless units such that $\Lambda=-3$. $F=dA$ is the $U(1)$ field strength tensor associated to the gauge field $A_M$ and $\phi$ and $\bar\phi$ are charged scalar fields dual to the operators $\mathcal{O}$ and $\bar{\mathcal{O}}$, respectively. The mass is chosen to be $m^2=-2$ so that the scalar operators are relevant and have conformal dimension $\Delta=2$. The action of the $U(1)$ transformation is given by
\begin{equation}\label{eq:bulk U(1) transformation}
\phi\rightarrow e^{-i q \alpha}\phi\,,\qquad \bar{\phi}\rightarrow e^{i q \alpha}\bar{\phi}\,,\qquad A_M\rightarrow A_M-\partial_M\alpha\,,
\end{equation}
and the covariant derivatives are defined as
\begin{equation}
\mathcal{D}_M\phi=\partial_M\phi-i q A_M\phi\,,\qquad\mathcal{D}_M\bar{\phi}=\partial_M\bar{\phi}+i q A_M\bar\phi\,,
\end{equation}
where $q$ is the charge, which we fix to unity. In the gravitational dual, the bulk action is Hermitian and the non-Hermiticity enters through the boundary conditions of the field $\phi$ and $\bar\phi$, thus mimicking how in the QFT we use the sources of $\mathcal{O}$ and $\bar{\mathcal{O}}$ to introduce the non-Hermiticty. 

In this paper, we are interested in studying the black hole interior of isotropic non-Hermitian configurations with finite temperature and no charge density.
To this end we consider the following ansatz 
\begin{align}\label{eq:Ansatz}
ds^2&=\frac{1}{z^2}\left(-g(z) dt^2+\frac{dz^2}{f(z)}+d\vec{x}^2\right)\,,\qquad A=0\,,\nonumber\\
\phi&=(1-\eta)\,\psi(z) \,,\qquad \bar{\phi}=(1+\eta)\,\psi(z)\,,
\end{align}
The AdS boundary is located at $z=0$ and without loss of generality we impose that there exists an event horizon at $z=1$. Hence, at $z=0$ we impose 
\begin{equation}
g(0)=1\,,\qquad f(0)=1\,,\qquad \partial_z\psi(0)=M\,,
\end{equation}
and at $z=1$ we impose 
\begin{equation}
g(1)=0\,,\qquad f(1)=0\,,
\end{equation}
so that we indeed have an event horizon. 
Note that, with our ansatz, the boundary conditions fix the source of operator $\mathcal{O}$ to be $(1+\eta)M$ while the source of $\bar{\mathcal{O}}$ is $(1-\eta)M$ thus achieving the non-Hermitian construction of the action \eqref{eq:QFT NH action}. Moreover, by choosing the bulk action of $\mathcal{PT}$, inherited from the dual QFT, to be
 \begin{subequations}
\begin{align}
    y=(x,z)=(t,x^1,x^2,z)\xrightarrow{\mathcal{PT}}\mathcal{PT}y=(\mathcal{PT}x,z)=(-t,-x^1,x^2,z)\,,\\
    \phi(y)\xrightarrow{\mathcal{PT}}\phi(\mathcal{PT}y)\,,\qquad \bar{\phi}(y)\xrightarrow{\mathcal{PT}}\bar{\phi}(\mathcal{PT}y)\,,
\end{align}
\end{subequations}
we find that the bulk theory is also $\mathcal{PT}$-symmetric as the action and the boundary conditions do not break this symmetry. Similarly, the Dyson map \eqref{eq:DysonMapBdry} can also be realized in the bulk as a $U(1)$ gauge transformation \eqref{eq:bulk U(1) transformation} with complex parameter
\begin{equation}
\alpha=\frac{i}{2q}\log\left(\frac{1-\eta}{1+\eta}\right)\,.
\end{equation}

The equations of motion of this model are given by
\begin{align}\label{eq:eoms}
\psi ''+ \left(\frac{f'}{2f}+\frac{g'}{2g}-\frac{2}{z}\right)\psi '-\frac{m^2\psi+2v (1-\eta^2)\,\psi^3}{f z^2}=0\\
f'+\frac{3}{z}(1-f)-\left(1-\eta ^2\right)\left[\frac{z}{2} f  \left(\psi '\right)^2+\frac{m^2}{2z}\psi^2+\frac{v}{2z}(1-\eta^2)\,\psi^4\right]=0\\
\frac{g'}{g}+\frac{3 }{z f} (1-f)+\left(1-\eta ^2\right)\left[\frac{z}{2} 
   \left(\psi '\right)^2-\frac{m^2}{2zf}\psi ^2-\frac{v}{2zf}(1-\eta^2)\,\psi^4\right]=0
\end{align}

To conclude this section we summarize here the relevant holographic dictionary needed to analyze the dual CFT. An explicit derivation can be found in appendix \ref{appendix:Holographic Dictionary}. Using the following asymptotic expansion for the fields
\begin{subequations}\label{eq:Asymptotic_Expansion}
\begin{align}
    g(z)&=1-g_3\, z^3+...\,,\\
    f(z)&=1+\frac{1}{2}M^2(1-\eta^2)z^2-\frac{1}{3}\left(3g_3-4(1-\eta^2)M\psi_v\right)z^3+...\,,\\
    \psi(z)&= Mz+\psi_v z^2+ ...\,,
\end{align}
\end{subequations}
obtained from solving the equations of motion for $z\rightarrow0$; we conclude that the vacuum expectation values (VEVs) of the operators $\mathcal{O}$ and $\bar{\mathcal{O}}$, and the boundary energy-momentum density; are given by 
\begin{equation}\label{eq:VEVs}
\expval{\mathcal{O}}=(1-\eta)\,\psi_v\,,\qquad \expval{\bar{\mathcal{O}}}=(1+\eta)\,\psi_v\,,\qquad \expval{T_{tt}}=2g_3-\frac{2}{3}(1-\eta^2)M\psi_v\,.
\end{equation}

\subsection{Phase diagram}

\begin{figure}
\centering
\begin{subfigure}{.49\textwidth}
    \includegraphics[width=\textwidth]{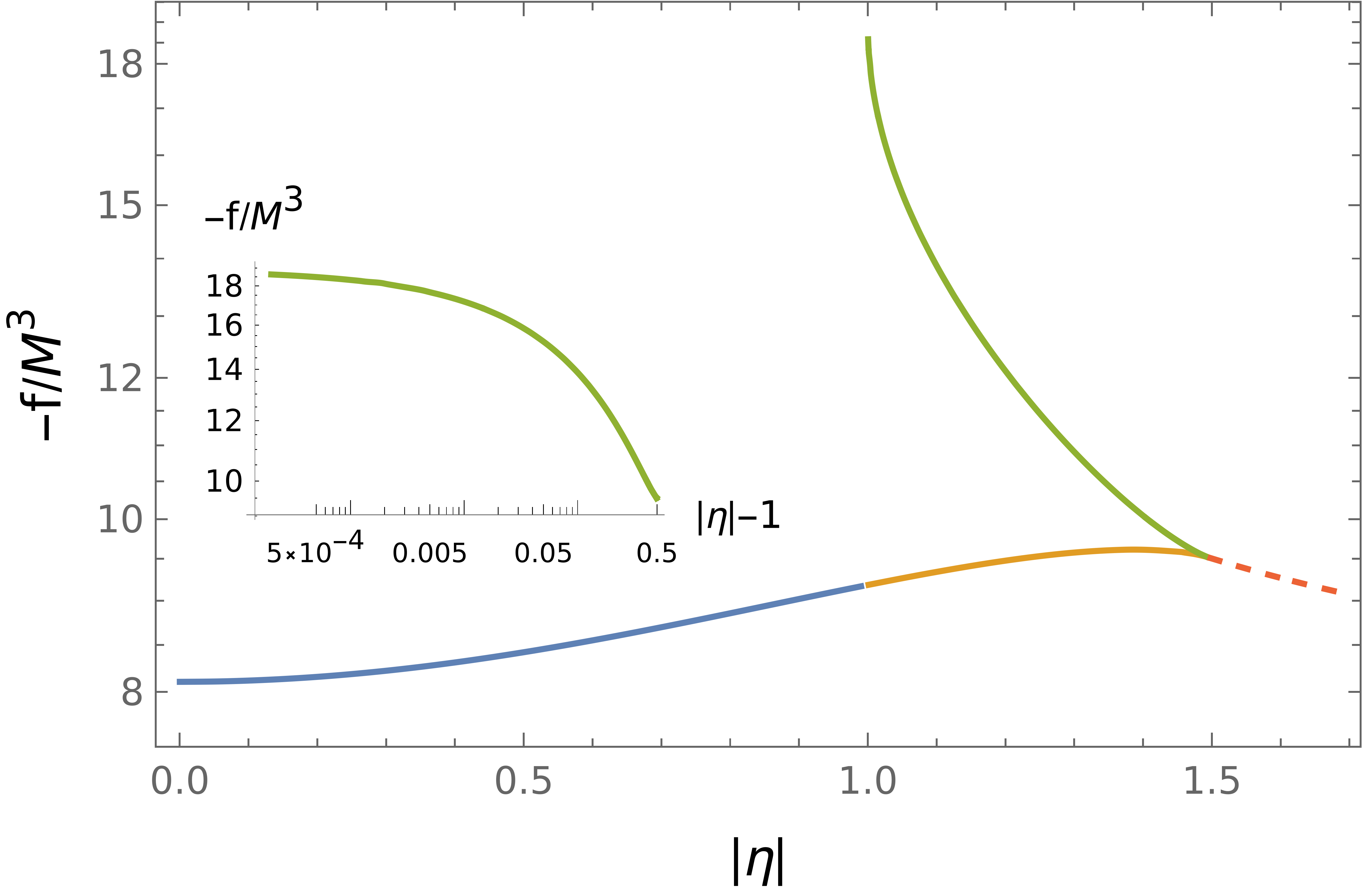}
    \caption{$T/M=0.5$}
    \label{fig:Current_FEplot_t05}
\end{subfigure}
\hfill
\begin{subfigure}{.5\textwidth}
    \includegraphics[width=\textwidth]{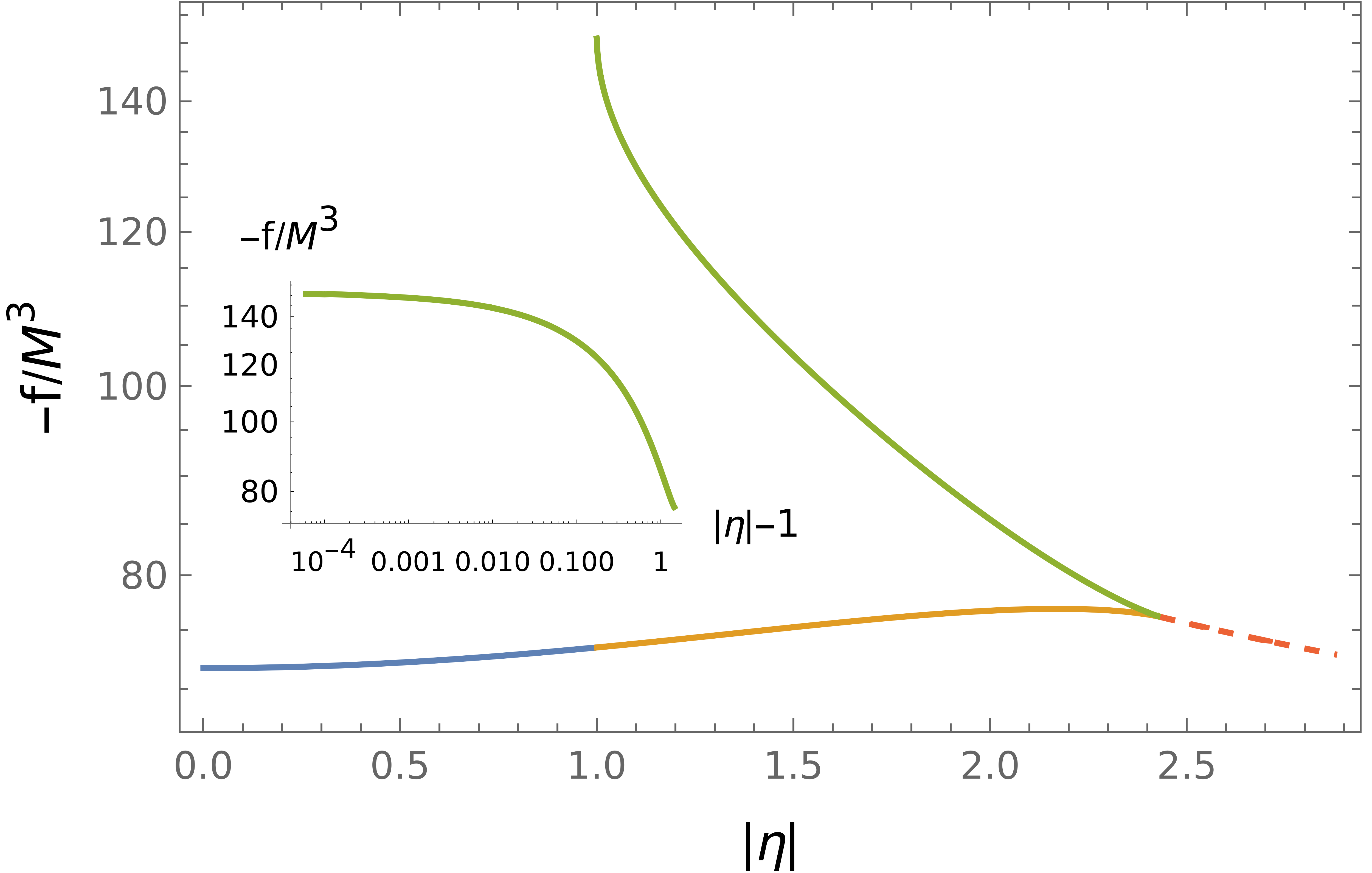}
    \caption{$T/M=1$}
    \label{fig:FEplot_t1}
\end{subfigure}
\caption{Free energy of solutions with $v=1$ and $T/M=0.5$ (left) and $T/M=1$ (right) as a function of $\eta$. For $|\eta|>1$ two branches of solutions exist. The green branch has lower free energy and is thus thermodynamically preferred. The two branches meet at $|\eta_c|\approx 1.49$ (left) and $|\eta_c|\approx 2.43$ (right). For $|\eta|>|\eta_c|$ only complex geometries exist: the red dashed line shows the real part of their free energy. In the insets we plot the free energy of the green branch in a log-log plot: it approaches a constant as $|\eta|\to 1^+$.}
\label{fig:FEplot}
\end{figure}

Following \cite{Xian:2023zgu}, here we summarize the phase diagram of the holographic model introduced in the previous section. In figure \ref{fig:FEplot} we plot the free energy \eqref{eq:free energy holo dic}
for $v=1$ and $|T|/M=\{0.5,1\}$.
We find three distinct phases.
\begin{itemize}
\item \textbf{$\boldsymbol{\mathcal{PT}}$-symmetric phase} ($|\eta|<1$): In this phase the NEC is preserved, the geometry is real and $\mathcal{PT}$ symmetry is not spontaneously broken. These solutions
(blue line in figure~\ref{fig:FEplot})
can be mapped to those of an equivalent Hermitian theory via the Dyson map \eqref{eq:DysonMapBdry}.

\item \textbf{$\boldsymbol{\mathcal{PT}}$-restored phase}  ($1<|\eta|<\eta_c$): In this phase the NEC is violated but the geometry is real and $\mathcal{PT}$ symmetry is not spontaneously broken. 
These solutions only exist at non-zero temperature and thus we identify them as the holographic duals of the $\mathcal{PT}$-restored solutions introduced in the toy model of section \ref{sec:TFD toy model}. We observe two branches: a subdominant one smoothly connected to the $\mathbf{\mathcal{PT}}$-symmetric phase
(yellow line in figure~\ref{fig:FEplot})
, and a dominant one disconnected from the $\mathbf{\mathcal{PT}}$-symmetric phase
(green line in figure~\ref{fig:FEplot}).
In both branches we cannot map the solutions to solutions of a Hermitian theory using the Dyson map \eqref{eq:DysonMapBdry}. 

\item \textbf{$\boldsymbol{\mathcal{PT}}$-broken phase} ($|\eta|>\eta_c$): In this phase the NEC is violated, the geometry is complex and $\mathcal{PT}$ symmetry is spontaneously broken. These solutions
(red dashed line in figure~\ref{fig:FEplot})
have complex temperature and thus circumvent our analysis of section \ref{sec:TFD toy model}. 

\end{itemize}

\section{Black hole interiors}\label{sec:BHInteriors}
In this section we discuss the features of the black hole interiors. We divide our discussion according to the different phases discussed in the previous section.  

\subsection{$\mathcal{PT}$-symmetric phase}
The $\mathcal{PT}$-symmetric phase can be mapped via a complexified $U(1)$ transformation to an equivalent Hermitian description with source $\tilde{M}=\sqrt{1-\eta^2} M$ and with the same geometry. Hence, the geometry of the black hole interior is the same as the one observed in \cite{Frenkel:2020ysx}. Near the singularity we have a Kasner cosmology of the form
\begin{equation}\label{eq:Kasner Cosmology}
ds^{2}=-d\tau^2+c_t\tau^{2p_t}dt^2+c_x\tau^{2p_x}d\vec{x}^2\,,\qquad \psi=-p_\psi \log(\tau)\,,
\end{equation}
where $\{c_t,c_x\}$ are constants, $\tau$ is a timelike coordinate defined as
\begin{equation}
\tau=\int \frac{dz}{z\sqrt{f}}\,,
\end{equation}
and the Kasner coefficients $\{p_t,p_x,p_\psi\}$ satisfy the following relations
\begin{equation}\label{eq:KasnerEq}
p_t+2p_x=1\,,\qquad p_t^2+2p_x^2+(1-\eta^2)p_\psi^2=1\,.
\end{equation}

\begin{figure}
\centering
\includegraphics[width=0.6\textwidth]{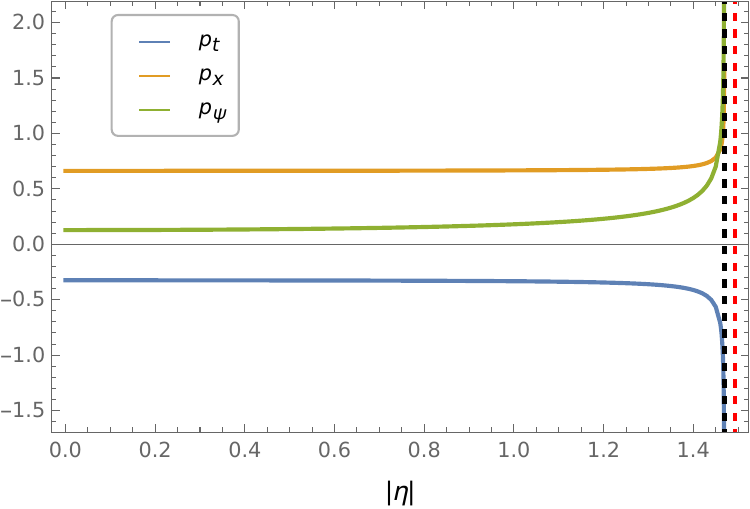}
\caption{Kasner coefficients for the non-Hermitian model with $v=1$ at $T/M=0.5$ in the $\mathcal{PT}$-symmetric phase ($|\eta|<1$) and in the subdominant branch of the $\mathcal{PT}$-restored phase ($1<|\eta|<\eta_c\approx 1.49$). The Kasner coefficients are continuous and diverge as $|\eta|\rightarrow\eta_s\approx 1.47$. For $\eta_s<|\eta|<\eta_c$ the subdominant branch of the $\mathcal{PT}$-restored phase does not have a Kasner singularity and instead has the same singularity as in the dominant branch. In the plot, the red dashed line corresponds to $\eta_c$ and the black dashed line to $\eta_s$.}
\label{fig:KasnerCoefsReal}
\end{figure}

We plot the Kasner coefficients as a function of $\eta$ in figure \ref{fig:KasnerCoefsReal}. As $|\eta|\rightarrow1$, the Kasner coefficients converge to those of the Schwarzschild-AdS black brane solution. This follows from the fact that at the exceptional point ($|\eta|=1$) the scalar field decouples from the Einstein equations and no longer backreacts on the geometry. 

\subsection{$\mathcal{PT}$-restored phase: Dominant branch}\label{subsec:PT restored dominant branch}
In the dominant branch of the $\mathcal{PT}$-restored phase, the black brane interior exhibits a much richer structure, completely different from that observed in the previous section.

Firstly, in the coordinates \eqref{eq:Ansatz}, we find that in the black hole interior the metric becomes non-analytic at $z=z_\sigma$. In particular we see that the fields admit an expansion in half-integer powers of $(z_\sigma-z)$ of the form 
\begin{align}\label{eq:zsigmaHitExpansion} 
g&=-g_0+\frac{g_0  \left(3  z_{\sigma }^2f_{3/2}+8(m^2\psi_0+2v\left(1-\eta^2\right)\psi_0^3) \sqrt{\left(\eta ^2-1\right) z_{\sigma }}\right)}{
   \left(6-m^2\left(1-\eta^2\right)\psi _0^2-v\left(1-\eta^2\right)^2\psi_0^4\right) z_{\sigma }}\left(z_{\sigma }-z\right)^{1/2}+...\,,\nonumber\\
f&=\frac{
   6-m^2\left(1-\eta^2\right)\psi _0^2-v\left(1-\eta^2\right)^2\psi_0^4 }{z_{\sigma }}\left(z_{\sigma
   }-z\right)+f_{3/2} \left(z_{\sigma }-z\right){}^{3/2}+...\,,\nonumber\\
\psi &=\psi _0-\frac{2 }{\sqrt{\left(\eta
   ^2-1\right) z_{\sigma }}}\left(z_{\sigma }-z\right)^{1/2}+...\,
\end{align}
We have checked numerically that our solutions agree with this expansion.

Despite this non-analyticity,
there is no curvature singularity at $z=z_\sigma$. We can see this by introducing $u=\sqrt{z_\sigma-z}$ as coordinate. In $u$-coordinates the metric is analytic and non-singular in the vicinity of $u=0$
\begin{equation}\label{eq:BounceMetric u}
ds^2=\frac{1}{\left(z_\sigma-u^2\right)^2}\left(\left(g_0+g_{1/2}\, u\right)dt^2-\frac{4u^2}{f_1 u^2+f_{3/2}\,u^3}du^2+d\vec{x}^2+...\right)\,,
\end{equation}
thus it follows that $u=0$ is not a singularity. 

Having established that the spacetime does not end at $z=z_\sigma$, we now discuss how to continue it beyond this point. One natural choice is to analytically continue the solution by assuming that the expansion \eqref{eq:zsigmaHitExpansion} continues to hold for $z>z_\sigma$. This would then imply that beyond $z_\sigma$ the black hole interior becomes complex. 

However, this perspective has two fundamental flaws. First, timelike and null geodesics cannot probe the complex region unless their affine parameter becomes complex. Hence, causal observers have to complexify as they cross
$z_\sigma$. Second, the standard interpretation of the eternal black hole as a TFD state with real temperature, consisting of two identical copies with the direction of time reversed, is lost. For spacelike geodesics we find that the usual relation between the imaginary part of the time-shift and the temperature no longer holds as the former now depends on the particular geodesic we are considering. As a consequence, we can no longer interpret the eternal black hole as describing the Schwinger-Keldysh contour associated to the TFD \cite{Herzog:2002pc,Skenderis:2008dh,Skenderis:2008dg}.
These observations suggest that the physically relevant continuation of the spacetime should be a different one. In particular, it should be one such that the black hole interior is always real. 

Guided by these ideas and noting that the metric \eqref{eq:BounceMetric u} can be continued to $u<0$ without any apparent issues, we propose that the physical continuation is instead to define a new patch to cover the region $u<0$. 
We choose to parametrize the $u<0$ region defining $u=-\sqrt{z_\sigma-\tilde{z}}$ with $\tilde{z}$ a new coordinate defined in $(0,z_\sigma)$. 
This choice is convenient as it allows us to describe the continuation from $u>0$ to $u<0$ as moving from one sheet of the solution to another one across the branch cut associated with $\sqrt{z_\sigma-z}$. Concretely, the solution for $\tilde{z}\rightarrow z_\sigma$ is obtained from \eqref{eq:zsigmaHitExpansion} by exchanging $\sqrt{z_\sigma-z}\rightarrow-\sqrt{z_\sigma-\tilde{z}}$
\begin{align}\label{eq:zsigmaBounceExpansion} 
g&=-g_0-\frac{g_0  \left(3  z_{\sigma }^2f_{3/2}+8(m^2\psi_0+2v\left(1-\eta^2\right)\psi_0^3) \sqrt{\left(\eta ^2-1\right) z_{\sigma }}\right)}{
   \left(6-m^2\left(1-\eta^2\right)\psi _0^2-v\left(1-\eta^2\right)^2\psi_0^4\right) z_{\sigma }}\left(z_{\sigma }-\tilde{z}\right)^{1/2}+...\,,\nonumber\\
f&=\frac{
   6-m^2\left(1-\eta^2\right)\psi _0^2-v\left(1-\eta^2\right)^2\psi_0^4 }{z_{\sigma }}\left(z_{\sigma
   }-\tilde{z}\right)-f_{3/2} \left(z_{\sigma }-\tilde{z}\right){}^{3/2}+...\,,\nonumber\\
\psi &=\psi _0+\frac{2 }{\sqrt{\left(\eta
   ^2-1\right) z_{\sigma }}}\left(z_{\sigma }-\tilde{z}\right)^{1/2}+... \,
\end{align}
We stress that the metric in $u$ coordinates is analytic at $u=0$ and, consequently, there is no jump in the extrinsic curvature at the surface $z=z_\sigma$. Hence when gluing the $z$-patch and $\tilde{z}$-patch we do not need to add a matter shell.

\begin{figure}
\centering
\begin{subfigure}{.485\textwidth}
    \includegraphics[height=0.63\textwidth]{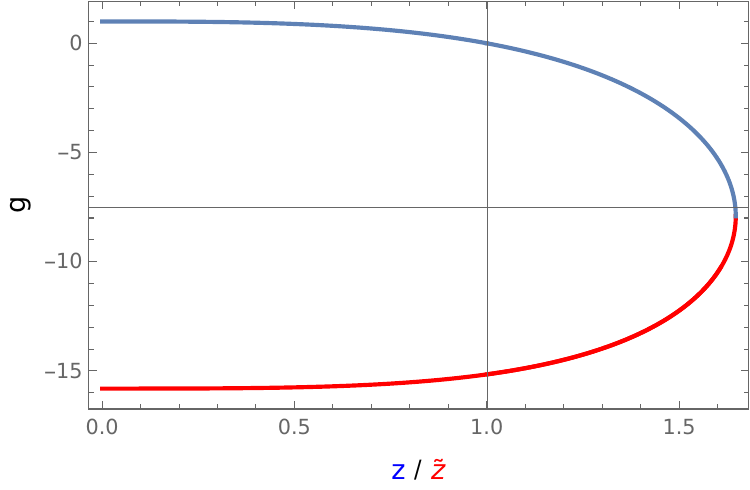}
    \caption{$g$}
    \label{fig:gBounce}
\end{subfigure}
\hfill
\begin{subfigure}{.495\textwidth}
    \includegraphics[height=0.63\textwidth]{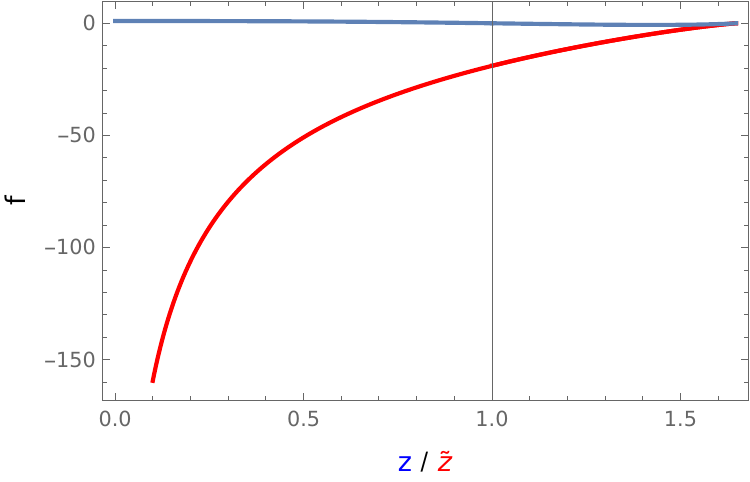}
    \caption{$f$}
    \label{fig:fBounce}
\end{subfigure}
\begin{subfigure}{.485\textwidth}
    \includegraphics[height=0.63\textwidth]{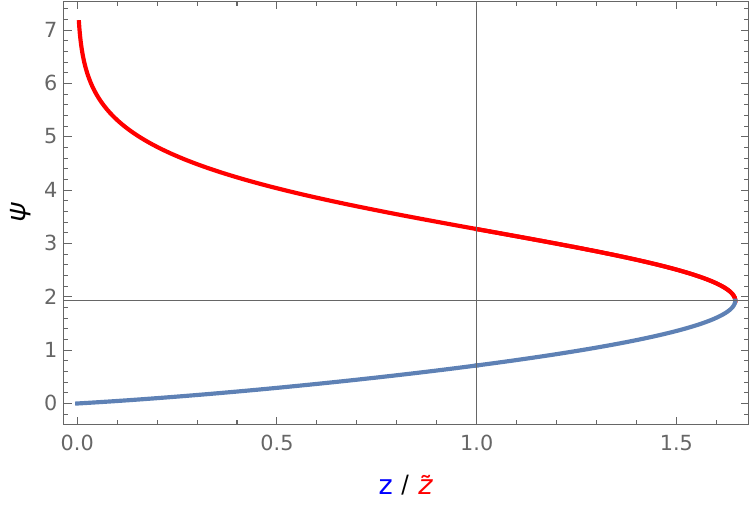}
    \caption{$\psi$}
    \label{fig:psiBounce}
\end{subfigure}
\hfill
\begin{subfigure}{.495\textwidth}
    \includegraphics[height=0.63\textwidth]{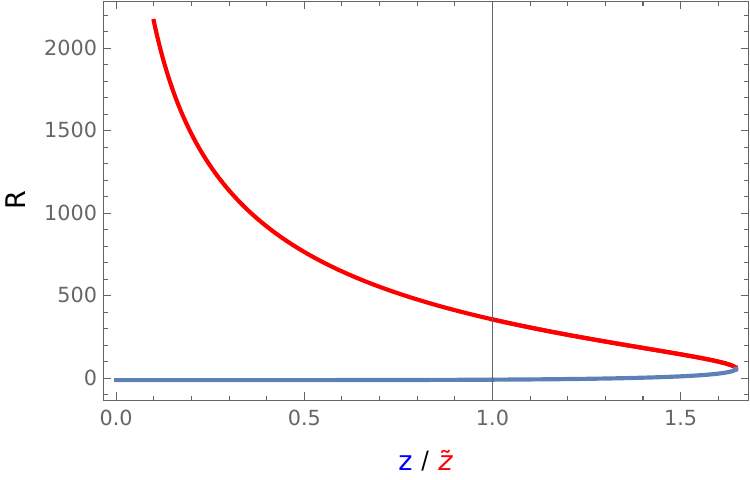}
    \caption{$R$}
    \label{fig:RBounce}
\end{subfigure}
\caption{Fields and Ricci scalar in the dominant branch of the $\mathcal{PT}$-restored phase for a solution with $v=1$, $T/M=0.5$ and $\eta=1.45$. The blue line represents the patch with coordinate $z$ while the red line denotes the patch with coordinate $\tilde{z}$. As $\tilde{z}\rightarrow0$, $g$ goes to a constant but $f$ and $\psi$ diverge. We conclude that $\tilde{z}\rightarrow0$ is a singularity as $R$ diverges as we approach that point.
}
\label{fig:BounceFieldsPlots}
\end{figure}

We plot a representative solution in figure \ref{fig:BounceFieldsPlots}. We can see that as $\tilde{z}\rightarrow0$, $g$ tends to a constant while $f$ and $\psi$ diverge. Moreover, we see that $\tilde{z}=0$ is a singularity as the Ricci diverges (figure \ref{fig:RBounce}). Remarkably, this singularity is not of Kasner type as it cannot be casted into the form \eqref{eq:Kasner Cosmology}. We can see this by noting that, as $g$ tends to a constant, $g_{tt}\propto g_{xx}$ and thus if we were to have a Kasner singularity
it would have $p_t=p_x$. Then from the Kasner relations \eqref{eq:KasnerEq} it would follow that
\begin{equation}
p_t=1/3\,,\qquad p_\psi^2=\frac{2}{3(1-\eta^2)}\,,
\end{equation}
and as $\eta^2>1$, this would imply that $p_\psi\in i\mathbb{R}$ which in turn would mean that $\psi$ must be complex. Since our numerical solution has $\psi$ real, 
we find a contradiction and thus our singularity cannot be of Kasner type.
We postpone a detailed discussion of the nature of this singularity to section \ref{sec:NewSingularity}. 

\subsection{$\mathcal{PT}$-restored phase: Subdominant branch}

The subdominant branch of the $\mathcal{PT}$-restored phase showcases two very distinct behaviors depending on the value of $|\eta|$. For $|\eta|<\eta_s$ we observe features
qualitatively similar  to those of the $\mathcal{PT}$-symmetric phase. Although the NEC is violated, the singularity is still of Kasner type. The corresponding Kasner coefficients satisfy the relation \eqref{eq:KasnerEq} and are continuously connected to those of the $\mathcal{PT}$-symmetric branch as illustrated in figure \ref{fig:KasnerCoefsReal}.
As $|\eta|$ approaches $\eta_s$, the Kasner coefficients diverge indicating the end of the Kasner behavior. 

For $|\eta|>\eta_s$ we instead observe 
the same qualitative behavior as in the dominant branch. We find that the metric has the same non-analytic structure as in \eqref{eq:zsigmaHitExpansion}, and performing the analytic continuation described in the previous section we again encounter a non-Kasner singularity.

\subsection{$\mathcal{PT}$-broken phase}
\begin{figure}
\centering
\includegraphics[width=0.6\textwidth]{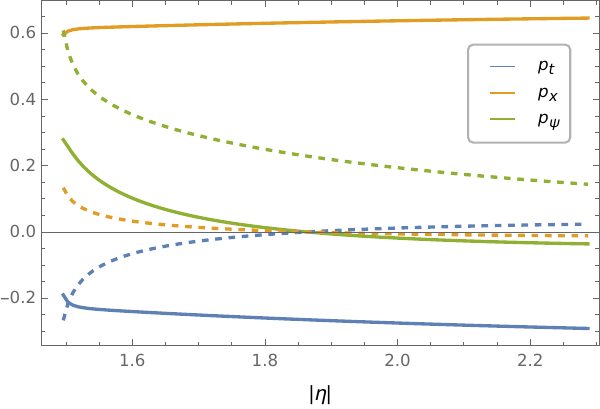}
\caption{Kasner coefficients for the non-Hermitian model with $v=1$ at $T/M=0.5$ in the $\mathcal{PT}$-broken phase. The solid lines denote the real part of the Kasner coefficients and the dashed line the imaginary part. In this phase the geometry is complex and so are the Kasner coefficients. }
\label{fig:KasnerCoefsComplex}
\end{figure}

In the $\mathcal{PT}$-broken phase the exterior geometry becomes complex and, consequently, so does the black hole interior. Nevertheless, we can still define Kasner coefficients following the procedure outlined in appendix \ref{app: Numerical Kasners}. Doing so, we find that despite the complex geometry we still have a Kasner singularity with complex Kasner coefficients satisfying the relations \eqref{eq:KasnerEq}. We plot the value of these coefficients in figure \ref{fig:KasnerCoefsComplex}. 
Note that although $\mathcal{PT}$ symmetry is spontaneously broken, the theory is $\mathcal{PT}$-symmetric and thus for each complex solution, the complex conjugate is also a solution \cite{Bender:2007nj, Fring:2022tll}. Hence together with the results presented here there is also a complex conjugate solution with Kasner exponents conjugate to the ones plotted in figure \ref{fig:KasnerCoefsComplex}.

\section{The new singularity}\label{sec:NewSingularity}

In this section we characterize the new singularity found for $|\eta|>1$. Notably the behavior of the singularity depends on the form of the scalar potential and we obtain qualitatively different results for $v=0$ and for $v\neq0$. In this section we collect the results for $v\neq0$ and we leave the case with $v=0$ to appendix~\ref{appendix:v0singularity}.

\subsection{Analytic form of the singularity}

\begin{figure}
\centering
\begin{subfigure}{.485\textwidth}
    \includegraphics[height=0.63\textwidth]{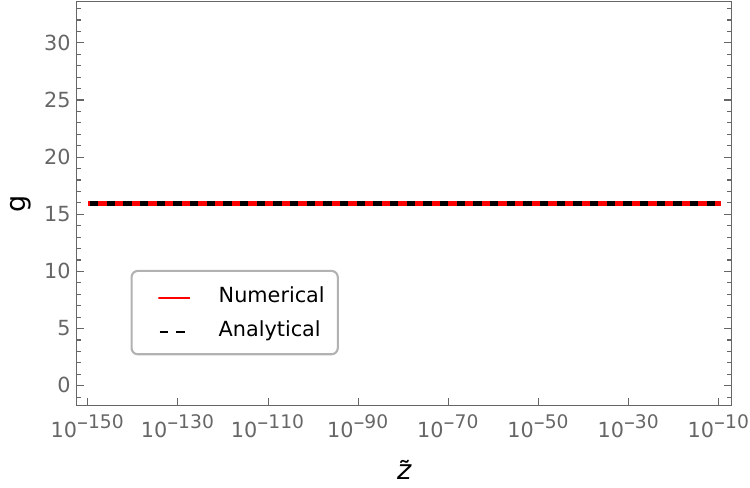}
    \caption{$g$}
    \label{fig:gSing}
\end{subfigure}
\hfill
\begin{subfigure}{.495\textwidth}
    \includegraphics[height=0.63\textwidth]{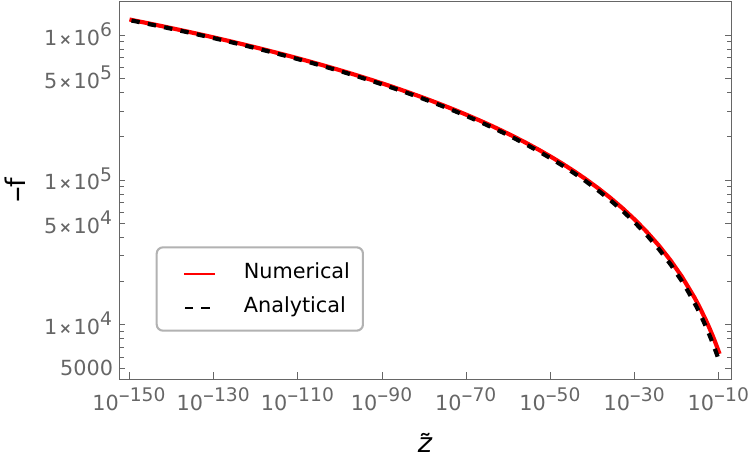}
    \caption{$f$}
    \label{fig:fSing}
\end{subfigure}
\begin{subfigure}{.485\textwidth}
    \includegraphics[height=0.63\textwidth]{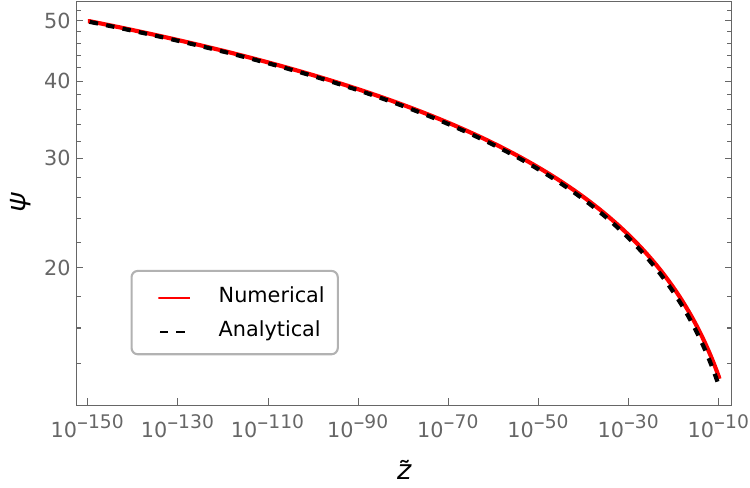}
    \caption{$\psi$}
    \label{fig:psiSing}
\end{subfigure}
\hfill
\begin{subfigure}{.495\textwidth}
    \includegraphics[height=0.63\textwidth]{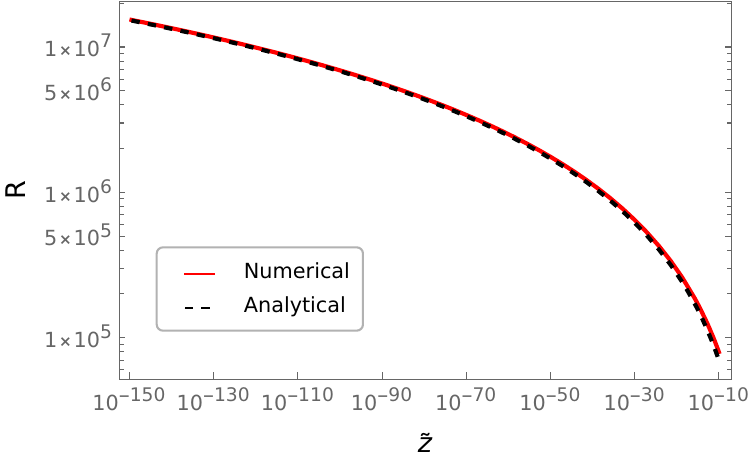}
    \caption{$R$}
    \label{fig:RSing}
\end{subfigure}
\caption{Fields and Ricci scalar near the non-Kasner singularity in the dominant branch of the $\mathcal{PT}$-restored phase for a solution with $v=1$, $T/M=0.5$ and $\eta=1.45$. The red lines represent the numerical result while the dashed black lines denote the analytical prediction. We see good agreement between analytical and numerical results.}
\label{fig:SingFieldsPlots}
\end{figure}

In figure \ref{fig:SingFieldsPlots}, we plot the behavior of the fields $\{g,f,\psi\}$ and of $R$ in the vicinity of $\tilde{z}=0$. As indicated in section \ref{subsec:PT restored dominant branch} we find that $g$ tends to a constant while both $f$ and $\psi$ diverge. We 
characterize the singularity by noting that our model admits the following approximate solution for $\tilde{z}\rightarrow0$
\begin{align}\label{eq:ExpansionNearSingularity}
g&\approx -g_0+...\,,\\
f&\approx -\frac{32}{3} v \left(\log\tilde{z}\right)^2-\frac{4}{9}(3m^2+8 v)\log \tilde{z}+\frac{4}{9}(3m^2+8 v)\log\left(-\log \tilde{z}\right) +...\,,\\
\psi &\approx \sqrt{-\frac{8}{\eta^2-1} \log \tilde{z} + \frac{3m^2+8v}{6v(\eta^2-1)}\log\left(-\log \tilde{z}\right)+...}\,,
\end{align}
which, as shown in figure \ref{fig:SingFieldsPlots}, agrees nicely with our numerical results. 
Importantly, the divergent behavior in $f$ and $\psi$ only depends on $m^2$, $v$ and $\eta$ and the only free coefficient that we find at this order is $g_0$, which can be reabsorbed into the definition of time. This has the important consequence that the leading near-singularity behavior of $R$, which we plot in figure \ref{fig:RSing}, is fully fixed in terms of the parameters $m^2$, $v$ and $\eta$ to be
\begin{equation}
    R=128 v \left(\log \tilde{z}\right)^2-\frac{16}{3}(4v-3m^2)\log \tilde{z}-\frac{16}{3}(3m^2+8v)\log \tilde{z} \log \left(-\log \tilde{z}\right)+...\,.
\end{equation}

To compare this singularity to Kasner, we introduce the comoving coordinate $\tau=\int\frac{dz}{z\sqrt{f}}$ to write the metric as
\begin{equation}\label{eq:NearSingularityMetric Comoving}
ds^2=-d\tau^2+\exp(2e^{\alpha\tau})\left(g_0dt^2+d\vec{x}^2\right)
\end{equation}
with $\alpha=4\sqrt{2v/3}$. In this form we can now meaningfully compare this result to the metric \eqref{eq:Kasner Cosmology}. Firstly, we find that while in Kasner the singularity takes place at the origin of the comoving coordinate, in our setup we have a singularity at $\tau\rightarrow\infty$. Moreover, our singularity is generically isotropic, which is not always the case for Kasner. 
Lastly, we can also compare the nature of the singularities by examining the Ricci scalars. In Kasner we find that it diverges as $1/\tau^2$, corresponding to a singularity happening at the origin of time (which is reached in finite proper time). On the other hand, in our geometry we find that the Ricci behaves as
\begin{equation}
R=6\alpha^2 \left(2e^{2\alpha\tau}+e^{\alpha\tau}\right)\,.
\end{equation}
Hence, we observe a singularity at $\tau \rightarrow \infty$, which takes infinite proper time to reach. In this sense, the singularity is milder for our geometry as it is technically only an asymptotic curvature blow-up and the spacetime is geodesically complete.

\subsection{Cosmological interpretation of the singularity}
The near-singularity geometry \eqref{eq:NearSingularityMetric Comoving} takes the form of a spatially flat FLRW spacetime,
\begin{equation}
ds^2=-d\tau^2+a(\tau)^2\left(dx_0^2+d\vec{x}^2\right)
\end{equation}
with scale factor $a(\tau)=\exp(\exp(\alpha \tau))$. Such accelerated expansion gives rise to a Little Rip at $\tau\rightarrow\infty$. In fact, the scale factor we find corresponds to model 1 in \cite{Frampton:2011sp}.
Hence we can interpret the near-singularity geometry as a Little Rip cosmology, whose existence is allowed by the violation of the NEC. 

Given the FLRW form of the metric, it is natural to interpret the geometry in terms of an effective perfect fluid with energy-momentum tensor
\begin{equation}
    T_{\mu\nu}=(\epsilon+p)\, u_\mu u_\nu+p\,g_{\mu\nu}\,,
\end{equation}
where $u^\mu=\delta^\mu_\tau$ is the comoving four-velocity. Substituting the scale factor into the Friedmann equations, we find that the corresponding fluid has equation of state
\begin{equation}
p=-\epsilon-2\alpha \sqrt{\frac{\epsilon}{3}}\,.
\end{equation}
Importantly, we have that $p<-\epsilon$, which indicates that indeed the NEC is violated. The energy density can be expressed directly in terms of the scale factor as
\begin{equation}
\epsilon=\frac{3\alpha^2}{4}\left(\log a\right)^2\,.
\end{equation}
Thus, unlike more conventional FLRW cosmologies where the energy density scales as a power of $a$, the cosmology found here is characterized by an energy density that grows quadratically with $\log a$.

\subsection{Heavy operator geodesics}
\label{subsec:HeavyOpGeodesics}
Having discussed the geometric properties of the Little Rip singularity, we now turn to its imprints on observables in the dual field theory. In particular, we seek to determine whether this novel singularity leaves signatures that distinguish it from a standard Kasner singularity. To this end, we consider two-sided correlators of heavy operators in the TFD state. In the limit of large conformal dimension $\Delta$, these correlators can be computed using the geodesic approximation \cite{Balasubramanian:1999zv,Louko:2000tp},
\begin{equation}
\expval{\mathcal{O}(t_L)\mathcal{O}(t_R)}\sim e^{-\Delta \mathcal{L}_{\rm ren}}\,,
\end{equation}
where $\mathcal{L}_{\rm ren}$ is the renormalized length of a spacelike bulk geodesic connecting the two asymptotic boundaries. In the limit of large conserved energy $E$, these geodesics penetrate deep into the black hole interior and become sensitive to the near-singularity geometry \cite{Fidkowski:2003nf,Festuccia:2005pi}. Their behavior as $E\rightarrow\infty$ therefore provides a direct diagnostic of the singularity; see \cite{Frenkel:2020ysx,Carballo:2024hem} for analogous signatures of Kasner interiors.

With this objective in mind, we consider radial spacelike geodesic connecting the two asymptotic boundaries. The corresponding equations of motion are
\begin{equation}
    \frac{g}{z^2}\frac{dt}{d\lambda}=E\,,\qquad -\frac{g }{z^2} \left(\frac{dt}{d\lambda}\right)^2+\frac{1}{f z^2} \left(\frac{dz}{d\lambda}\right)^2=1\,,
\end{equation}
where $\lambda$ is the affine parameter and $E$ the energy. The renormalized geodesic length of these geodesics is given by 
\begin{equation}\label{eq:GeodesicLengthFull}
\mathcal{L}_{\rm ren}=2\int_\epsilon^{z_\sigma}\frac{dz}{z}\sqrt{\frac{g}{f(E^2 z^2+g)}}+2\int_{z_\sigma}^{\tilde{z}_*}\frac{d\tilde{z}}{\tilde{z}}\sqrt{\frac{g}{f(E^2 \tilde{z}^2+g)}}-2\log(\epsilon)\,,
\end{equation}
where we have
assumed that the geodesic extends into the second patch of the geometry, and we have introduced the turning point $\tilde{z}_*$ as the solution of $E^2=-g(\tilde{z}_*)/\tilde{z}_*^2$ in that patch.
As explained above, in the limit $E\to\infty$ the geodesic approaches a null trajectory and probes the singularity. Hence, the non-analytic terms in the large-$E$ expansion of the geodesic length encode information about the singularity. Following \cite{Frenkel:2020ysx},
these terms can be extracted by evaluating the geodesic length in the near-singularity metric,
\begin{equation}
\label{eq:LRmetric}
g\approx -g_0\,,\qquad f\approx -\alpha^2\left( \log\tilde{z}\right)^2=-\frac{32}{3}v \left( \log\tilde{z}\right)^2\,,
\end{equation}
and taking the $E\rightarrow\infty$ limit of that result at the end of the computation. Thus, the expression we need to evaluate is given by
\begin{equation}\label{eq: Lsing step0}
\mathcal{L}_\text{sing}(E)=\lim_{\tilde{z}_*\rightarrow0} \int^{\tilde{z}_*}_{\tilde{z}_{\text{LR}}} \frac{d\tilde{z}}{\alpha\,\tilde{z}\log\tilde{z}}\sqrt{\frac{1}{\tilde{z}^2/\tilde{z}_*^2-1}}\,,
\end{equation}
where $\tilde{z}_{\text{LR}}$ denotes a cutoff below which we consider the geometry to be given by the Little Rip cosmology~\eqref{eq:LRmetric}. 
Although this integral does not admit a closed form expression, we can infer its asymptotic behavior in the large-$E$ (small-$\tilde{z}_*$) limit. To this end, we introduce the rescaled integration variable $u=\tilde{z}/\tilde{z}_*$, in terms of which the integral \eqref{eq: Lsing step0} reads
\begin{align}\label{eq: Lren Final}
\mathcal{L}_\text{sing}(E)&=\lim_{\tilde{z}_*\rightarrow0} \int^{1}_{\tilde{z}_{\text{LR}}/\tilde{z}_*} \frac{du}{\alpha u\left(\log\tilde{z}_* +\log u\right)}\sqrt{\frac{1}{u^2-1}}\nonumber \\
&=\lim_{\tilde{z}_*\rightarrow0}\frac{1}{\alpha\log\tilde{z}_*}\int^{1}_{\infty} \frac{du}{ u}\sqrt{\frac{1}{u^2-1}}=-\lim_{\tilde{z}_*\rightarrow0}\frac{\pi}{2\alpha\log\tilde{z}_*}\nonumber\\
&= \frac{\pi}{2\alpha \log\left(E/\sqrt{g_0}\right)}\,,
\end{align}
where in the last line we have substituted the definition of $\tilde{z}_*$ in terms of $E$.
Importantly, we find that while for Kasner the geodesic length goes to zero as a power law of $E$, in our case it 
vanishes more slowly than any power law and it behaves non-algebraically. 
This behavior is related to the asymptotic nature of the Little Rip. Intuitively, as the curvature blow-up is reached only at infinite cosmological proper time, the geodesic turning point approaches the singularity slower than for the Kasner singularity.
The Little Rip therefore leaves a qualitatively distinct imprint on heavy-operator correlators, allowing it to be distinguished from a standard Kasner interior.

Finally, we compute the geodesic time shift in the large-$E$ limit and find the following result
\begin{equation}
t_\text{bdry}=\int_0^{z_\sigma}\frac{dz}{g}\sqrt{\frac{gE^2}{f(E^2 z^2+g)}}+\int_{z_\sigma}^{\tilde{z}_*}\frac{d\tilde{z}}{g}\sqrt{\frac{gE^2}{f(E^2 \tilde{z}^2+g)}}=t_\text{sing}+\frac{1}{E}+...\,,
\end{equation}
where $t_\text{sing}$ is the time-shift for a null geodesic. The $1/E$ term is a non-analytic contribution associated with the near-boundary geometry, which dominates in the large-$E$ expansion over both the analytic terms as well as non-analytic contributions coming from the near-singularity region. 
Expressing the near-singularity contribution to the geodesic length \eqref{eq: Lren Final} in terms of $\Delta t=t_\text{bdry}-t_\text{sing}$, we obtain
\begin{equation}
\mathcal{L}_\text{sing}(\Delta t)= -\frac{\pi}{2\alpha \log\left(\sqrt{g_0}\,\Delta t\right)}\,.
\end{equation}
which again differs from the power-law behavior observed in Kasner as it decays to zero slower than any power law.

\section{Conclusions}\label{sec:Conclusions}

We have studied the black hole interiors of the holographic non-Hermitian, $\mathcal{PT}$-symmetric model of \cite{Arean:2019pom}. The $\mathcal{PT}$-restored and $\mathcal{PT}$-broken phases of this model violate the null energy condition (NEC) and exhibit novel phenomena associated with their non-Hermitian character. Of particular relevance here, NEC violation allows the interior dynamics to depart from the standard BKL paradigm, opening the possibility of qualitatively new near-singularity behavior.

In the $\mathcal{PT}$-restored phase, we identify regions of parameter space in which the interior approaches a novel non-Kasner asymptotic regime. Its detailed form depends sensitively on the scalar potential, leading to different scaling behaviors for quartic and quadratic potentials. In both cases, however, the deep-interior geometry becomes isotropic and can be interpreted as an FLRW cosmology approaching a Little Rip supported by NEC-violating matter. The curvature diverges only as the cosmological proper time tends to infinity, so the resulting spacetime is geodesically complete despite exhibiting an asymptotic curvature blow-up.

To determine whether this regime can be detected from the dual QFT, we studied two-sided correlators of heavy operators in the geodesic approximation. Although the relevant geodesic integrals do not admit closed-form expressions, we 
were able to solve them in the limit of large conserved energy $E$. Importantly we found that
the renormalized geodesic length vanishes as an inverse logarithm, thus more slowly than any power of $E$,
whereas Kasner geometries produce a power-law decay. The Little Rip interior therefore leaves a characteristic imprint on boundary correlators that allows it to be distinguished from the conventional Kasner regime.

These results provide a first step toward formulating a Little Rip/CFT correspondence. Our construction demonstrates that a Little Rip cosmology hidden behind a black hole horizon can nevertheless be encoded in observables of a non-Hermitian boundary theory. It would be interesting to understand how universal this encoding is, which additional QFT observables retain information about the Little Rip regime, and whether different asymptotic cosmologies parametrized by their scale factors can be organized into distinct classes from the boundary perspective.

Despite the violation of the NEC, we also find Kasner behavior for part of the $\mathcal{PT}$-restored phase and for all the solutions in the $\mathcal{PT}$-broken phase. The latter result is particularly striking because the bulk geometry is complex. Nevertheless, complex Kasner exponents can still be extracted and satisfy the usual Kasner relations. Thus, NEC violation does not necessarily eliminate Kasner behavior, but allows additional asymptotic regimes that are inaccessible to conventional NEC-preserving matter.

Several directions deserve further investigation. A natural extension would be to study charged non-Hermitian black holes and their interiors. Even in conventional Hermitian holographic models, finite charge density gives rise to a richer phase structure and qualitatively new interior dynamics \cite{Hartnoll:2020rwq,Carballo:2024hem}; its interplay with non-Hermiticity and $\mathcal{PT}$ symmetry may therefore lead to an even more intricate phase diagram. Within superconducting setups, it would also be interesting to determine whether the Josephson oscillations identified in \cite{Hartnoll:2020fhc} persist under non-Hermitian deformations and whether an analogous phenomenon accompanies the Little Rip regime. Another natural possibility is to probe the interior using suitable complexity=anything observables \cite{Jorstad:2023kmq,Arean:2024pzo}, which may provide an independent diagnostic of the departure from Kasner scaling. The thermal $a$-function \cite{Caceres:2022smh} is another potentially useful probe. Its standard monotonicity is tied to the NEC and, in the solutions studied here, the conventional quantity is no longer monotonically decreasing. Whether an analogous quantity exists and admits a meaningful interpretation as a measure of degrees of freedom in a non-Hermitian open-system setting remains an important question. Finally, the strong dependence of the Little Rip asymptotics on the scalar potential suggests that broader classes of non-Hermitian holographic models may realize an even richer landscape of exotic black hole interiors.

\section*{Acknowledgments}
We thank C. Hoyos and K. Landsteiner for useful and insightful discussions,
A. Serantes for his comments that led to the inclusion of the cosmological interpretation of the singularity, and H.-S. Jeong for comments on the draft. 
D.G.F. acknowledges the hospitality of Gent University and APCTP during the completion of this work.
This work is supported through the grants CEX2020-001007-S and PID2021-123017NB-I00 funded by MCIN/AEI/10.13039/501100011033 and by ERDF ``A way of making Europe''.
The work of D.G.F. is supported by FPI grant PRE2022-101810. J.F.P. is supported by the ‘Atracción de Talento’ program of the Comunidad de Madrid under grant 2020-T1/TIC-20495.

\appendix

\section{Holographic dictionary}\label{appendix:Holographic Dictionary}
In this appendix, we follow the standard holographic renormalization procedure \cite{Skenderis:2002wp}, to compute the one-point functions presented in equation \eqref{eq:VEVs}.

We commence by first defining the renormalized action $\mathcal{S}_{ren}$ associated with the bulk action \eqref{eq:Bulk_Grav_Action}
\begin{equation}\label{eq:Renormalized_Grav_Action}
    \mathcal{S}_{ren}=\int d^4y \sqrt{-g}\left(R-2\Lambda-\frac{1}{4}F_{MN}F^{MN}-\mathcal{D}_M\phi \mathcal{D}^M\bar{\phi}-m^2\phi\bar{\phi}-v\phi^2\bar{\phi}^2\right)+\mathcal{S}_{ct}\,,
\end{equation}
where $\mathcal{S}_{ct}$ is the standard counter-term (see e.g. \cite{Hartnoll:2008kx}) 
\begin{equation}\label{eq:Counterterm_Grav_Action}
    \mathcal{S}_{ct}=\int_{z=\epsilon} d^3x \sqrt{-\gamma}\left(2K-4-\phi\bar{\phi}\right)\,.
\end{equation}
Here, $\gamma_{\mu\nu}$ is the induced metric at the $z=\epsilon\rightarrow0$ hypersurface and $K$ is the trace of the extrinsic curvature $K_{\mu\nu}$.

Setting $\mathcal{S}_{ren}$ on-shell and using the asymptotic expansion \eqref{eq:Asymptotic_Expansion}, we find that the free energy is given by 
\begin{equation}\label{eq:free energy holo dic}
F=- V_2\left(g_3+\frac{2}{3}(1-\eta^2)M\psi_v \right)
\end{equation}
where $S_\text{ren,E}$ is the renormalized euclidean action and $V_2$ the volume of the boundary spatial manifold.

To compute the one-point functions we need the variation of the renormalized action 
\begin{align}\label{eq:delta_Renormalized_Grav_Action_NoSeries}
    \delta S_{ren}  =   &\int_{z=\epsilon} d^3x \sqrt{-\gamma } \left(K_{\mu \nu }-\left(K-2-\frac{1}{2} \phi  \bar{\phi }\right) \gamma _{\mu \nu }\right)\delta \gamma ^{\mu \nu } \nonumber \\
   &-\int_{z=\epsilon} d^3x \sqrt{-\gamma } \left(n_{\alpha } \left(\delta  \phi   \mathcal{D}^{\alpha }\bar{\phi }+\delta \bar{\phi} \mathcal{D}^{\alpha } \phi  +\delta A_{\mu } F^{\alpha \mu }\right)+  \bar{\phi }\delta
   \phi +\phi \delta  \bar{\phi }\right)\,,
\end{align}
where $n_\alpha$ is the outward-pointing unit normal vector corresponding to the $z=\epsilon$ boundary. Plugging in the ansatz \eqref{eq:Ansatz}, the asymptotic expansion \eqref{eq:Asymptotic_Expansion} and redefining the fluctuations as
\begin{equation}
    \delta\gamma^{\mu\nu}=\delta\gamma^{\mu\nu}_{(b)}\,z^{2}\,,\qquad  \delta A_\mu=\delta  A_\mu^{(b)} \,,\qquad  \delta\phi=\delta\phi^{(b)}\,z\,,\qquad \delta\bar{\phi}=\delta\bar{\phi}^{(b)}\,z\,,
\end{equation}
the above expression simplifies to
\begin{align}\label{eq:delta_Renormalized_Grav_Action_YesSeries}
    \delta S_{ren}  =&\int_{z=\epsilon} d^3x \left(-g_3+\frac{1}{3}(1-\eta^2)M\psi_v \right)\delta\gamma^{tt}_{(b)} \nonumber\\
   &+\int_{z=\epsilon} d^3x \left(-\frac{1}{2}g_3-\frac{1}{3}(1-\eta^2)M \psi_v \right)\left(\delta\gamma^{11}_{(b)}+\delta\gamma^{22}_{(b)}\right) \nonumber\\
   &+\int_{z=\epsilon} d^3x \left( (1+\eta)\psi_v \delta\phi^{(b)} +(1-\eta)\psi_v \delta\bar{\phi}^{(b)}\right)+ \text{terms of order $\epsilon$}
\end{align}
Finally, upon considering the usual holographic prescription for flat AdS boundaries 
\begin{equation}
    \expval{T_{\mu\nu}}=-2\lim_{\epsilon\rightarrow0}\frac{\delta \mathcal{S}_{ren}}{\delta \gamma^{\mu\nu}_{(b)} }\,,\quad
    \expval{\mathcal{O}}=\lim_{\epsilon\rightarrow0}\frac{\delta \mathcal{S}_{ren}}{\delta \bar{\phi}^{(b)} }\,,\quad \expval{\bar{\mathcal{O}}}=\lim_{\epsilon\rightarrow0}\frac{\delta \mathcal{S}_{ren}}{\delta \phi^{(b)} }\,,
\end{equation}
we recover the one-point functions presented in equation \eqref{eq:VEVs}.

\section{The new singularity for a quadratic potential}
\label{appendix:v0singularity}
In this appendix we discuss in detail the non-Kasner singularity in the absence of quartic self-interaction ($v=0$), i.e. when the potential is purely quadratic.

\subsection{Analytic form of the singularity}
\begin{figure}
\centering
\begin{subfigure}{.485\textwidth}
    \includegraphics[height=0.63\textwidth]{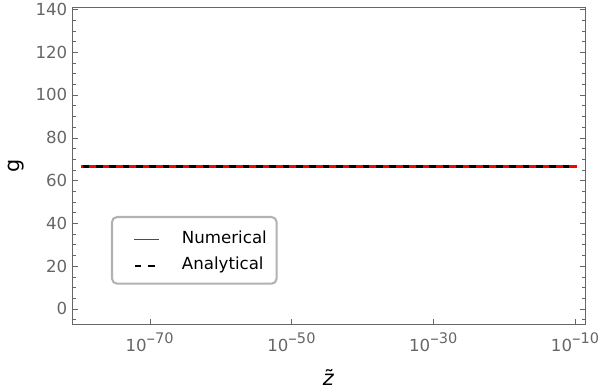}
    \caption{$g$}
    \label{fig:gSingv0}
\end{subfigure}
\hfill
\begin{subfigure}{.495\textwidth}
    \includegraphics[height=0.63\textwidth]{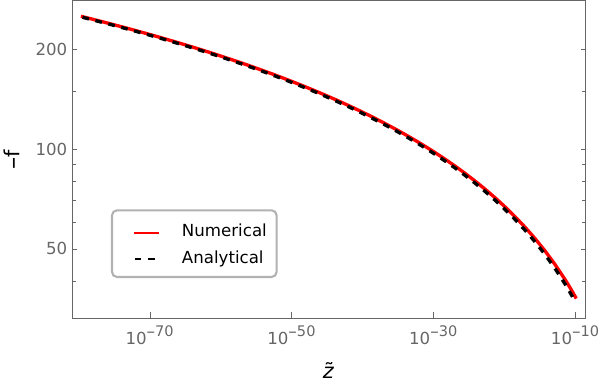}
    \caption{$f$}
    \label{fig:fSingv0}
\end{subfigure}
\begin{subfigure}{.485\textwidth}
    \includegraphics[height=0.63\textwidth]{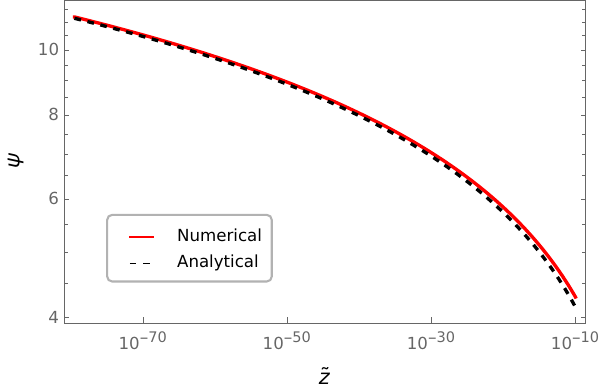}
    \caption{$\psi$}
    \label{fig:psiSingv0}
\end{subfigure}
\hfill
\begin{subfigure}{.495\textwidth}
    \includegraphics[height=0.63\textwidth]{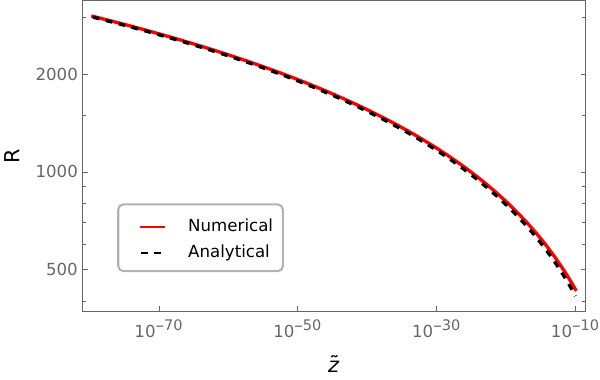}
    \caption{$R$}
    \label{fig:RSingv0}
\end{subfigure}
\caption{Fields and Ricci scalar near the non-Kasner singularity in the dominant branch of the $\mathcal{PT}$-restored phase for a solution with $v=0$, $T/M=0.5$ and $\eta=1.45$. 
The red lines represents the numerical result while the dashed black lines denote the analytical prediction. We see good agreement between analytical and numerical results.}
\label{fig:SingFieldsPlotsv0}
\end{figure}

In figure \ref{fig:SingFieldsPlotsv0}, we plot the behavior of the fields $\{g,f,\psi\}$ and of $R$ in the vicinity of $\tilde{z}=0$. As indicated in section \ref{subsec:PT restored dominant branch} we find that $g$ goes like a constant while both $f$ and $\psi$ diverge. This singularity corresponds to the following approximate solution for $\tilde{z}\rightarrow0$
\begin{align}\label{eq:ExpansionNearSingularityv0}
g&\approx -g_0+...\,,\\
f&\approx -\frac{2}{3} m^2 \log\tilde{z}+\frac{10}{9}(9-m^2)\log\left(-\log \tilde{z}\right) +...\,,\\
\psi &\approx \sqrt{-\frac{4}{\eta^2-1} \log \tilde{z} + \frac{2(9-m^2)}{3m^2(\eta^2-1)}\log\left(-\log \tilde{z}\right)+...}\,,
\end{align}
which, as shown in figure \ref{fig:SingFieldsPlotsv0}, agrees nicely with our numerical results. 
As for the solutions with quartic potential, the leading near-singularity behavior of $R$, which we plot in figure \ref{fig:RSingv0}, is fully fixed in terms of the parameters of the model $m^2$ and $\eta$ to be
\begin{equation}
    R=8m^2\log \tilde{z}+\frac{4}{3}(9-m^2)\log \left(-\log \tilde{z}\right)+...\,.
\end{equation}

Introducing the comoving coordinate $\tau=\int\frac{dz}{z\sqrt{f}}$ we can write metric as
\begin{equation}\label{eq:NearSingularityMetric Comoving v0}
ds^2=-d\tau^2+\exp(2\alpha\tau^2)\left(g_0dt^2+d\vec{x}^2\right)
\end{equation}
with $\alpha=|m^2|/6$. As for the solutions with quartic potential, we find that we have an isotropic singularity at $\tau\rightarrow\infty$. In this case however, the divergence is a power law as in Kasner
\begin{equation}
R=12\alpha (1+4\alpha \tau^2)
\end{equation}

\subsection{Cosmological interpretation of the singularity}
The near-singularity geometry \eqref{eq:NearSingularityMetric Comoving v0} takes the form of a spatially flat FLRW spacetime,
\begin{equation}
ds^2=-d\tau^2+a(\tau)^2\left(dx_0^2+d\vec{x}^2\right)
\end{equation}
with scale factor $a(\tau)=\exp(\alpha \tau^2)$. Such accelerated expansion gives rise to a Little Rip at $\tau\rightarrow\infty$.
Hence as before, we can interpret the near-singularity geometry as a Little Rip cosmology, whose existence is allowed by the violation of the NEC. 

Given the FLRW form of the metric, it is natural to interpret the geometry in terms of an effective perfect fluid with energy-momentum tensor
\begin{equation}
    T_{\mu\nu}=(\epsilon+p) u_\mu u_\nu+p g_{\mu\nu}\,,
\end{equation}
where $u^\mu=\delta^\mu_\tau$ is the comoving four-velocity. Substituting the scale factor into the Friedmann equations, we find that the corresponding fluid has equation of state
\begin{equation}
p=-\epsilon-2\alpha \,.
\end{equation}
Importantly, we have that $p<-\epsilon$, which indicates that indeed the NEC is violated. The energy density can be expressed directly in terms of the scale factor as
\begin{equation}
\epsilon=3\alpha \log a\,.
\end{equation}
Thus, unlike more conventional FLRW cosmologies where the energy density scales as a power of $a$, the cosmology found here is characterized by an energy density that grows with $\log a$.

\subsection{Heavy operator geodesics}

Following the same procedure as outlined in section \ref{subsec:HeavyOpGeodesics} we find that the non-analytic terms in the large energy expansion coming from the singularity can be obtained by evaluating the geodesic length in the near-singularity geometry. Using that
\begin{equation}
g\approx -g_0\,,\qquad f\approx f_0\log(z)=-\frac{2m^2}{3} \log(z)
\end{equation}
we thus conclude that we have to compute
\begin{equation}
\mathcal{L}_\text{sing}(E)=-\lim_{z_*\rightarrow\infty}\int^{z_*}_{z_{\text{LR}}}\frac{dz}{z}\sqrt{\frac{1}{-f_0 \log(z)(z^2/z_*^2-1)}} 
\end{equation}
As for the case with quartic potential, this integral does not admit a closed form expression but we can infer its large-$E$ (small $\tilde{z}_*$) behavior introducing the rescaled integration variable $u=\tilde{z}/\tilde{z}_*$
\begin{align}\label{eq: Lrenv0 Final}
\mathcal{L}_\text{sing}(E)&=-\lim_{\tilde{z}_*\rightarrow0} \int^{1}_{\tilde{z}_{\text{LR}}/\tilde{z}_*} \frac{du}{ u\sqrt{-f_0\log\tilde{z}_* -f_0\log u}}\sqrt{\frac{1}{u^2-1}}\nonumber \\
&=-\lim_{\tilde{z}_*\rightarrow0}\frac{1}{\sqrt{-f_0\log\tilde{z}_*}}\int^{1}_{\infty} \frac{du}{ u}\sqrt{\frac{1}{u^2-1}}=\lim_{\tilde{z}_*\rightarrow0}\frac{\pi}{2\sqrt{-f_0\log\tilde{z}_*}}\nonumber\\
&= \frac{\pi}{2\sqrt{f_0 \log\left(E/\sqrt{g_0}\right)}}\,.
\end{align}
where in the last line we have substituted the definition of $\tilde{z}_*$ in terms of $E$. We see that, as in the presence of the quartic term, the geodesic length vanishes slower than any power law and thus yields a qualitatively different imprint from the imprint associated with the Kasner singularity.

Finally, we compute the geodesic time shift in the large-$E$ limit and find the following result
\begin{equation}
t_\text{bdry}=\int_0^{z_\sigma}\frac{dz}{g}\sqrt{\frac{gE^2}{f(E^2 z^2+g)}}+\int_{z_\sigma}^{\tilde{z}_*}\frac{d\tilde{z}}{g}\sqrt{\frac{gE^2}{f(E^2 \tilde{z}^2+g)}}=t_\text{sing}+\frac{1}{E}+...\,,
\end{equation}
where $t_\text{sing}$ is the time-shift for a null geodesic. As in section \ref{subsec:HeavyOpGeodesics}, the $1/E$ term is a non-analytic contribution associated with the near-boundary geometry, which dominates in the large-$E$ expansion over both the analytic terms as well as non-analytic contributions coming from the near-singularity region. 
Expressing the near-singularity contribution to the geodesic length \eqref{eq: Lrenv0 Final} in terms of $\Delta t=t_\text{bdry}-t_\text{sing}$, we obtain
\begin{equation}
\mathcal{L}_\text{sing}(\Delta t)=\frac{\pi}{2\sqrt{-f_0 \log\left(\sqrt{g_0}\Delta t\right)}}\,.
\end{equation}
which again differs from the power-law behavior observed in Kasner as it decays to zero slower than any power law.

\section{Numerical methods}\label{app: Numerical methods}
In this appendix we discuss the numerical methods employed throughout the text. We perform our computations using Mathematica and working with $15\times$MachinePrecision.

To compute the field configuration outside the black hole, we discretize the coordinate $z$ in a Chebyshev-Gauss-Lobatto grid and we use Chebyshev derivative matrices to represent the derivative operators in the grid of 100 points. We fix boundary conditions according to the discussion of section \ref{sec:Holographic model} and we solve using a Newton-Raphson algorithm, which we stop when the fields change less than $10^{-50}$ in a step of the Newton-Raphson.

Near the event horizon ($z=1$), the fields satisfy the following expansion 
\begin{align}
f&=-\frac{1}{2}\left(6-m^2(1-x^2)\psi_h^2-v\left(1-\eta^2\right)^2\psi_h^4 \right)(z-1)+...\,,\\
g&=g_h(z-1)+...\,,\\
\psi&=\psi_h-\frac{2m^2\psi_h+4v\left(1-\eta^2\right)\psi_h^3}{6-m^2(1-\eta^2)\psi_h^2-v(1-\eta^2)^2\psi_0^4}(z-1)+...\,
\end{align}
where $\psi_h$ and $g_h$ are two free coefficients that can be read from the data of the exterior solution. Knowing $\psi_h$ and $g_h$, we obtain the field configuration in the black hole interior by evolving the data from $z=1+10^{-20}$ to deep in the black hole interior using NDSolve. 

For the solutions with Kasner singularity, we compute the interior until we can unequivocally identify the Kasner behavior. 
For the solutions without Kasner singularity, we compute the interior until we find a point where the numerics break down due to the square root behavior associated with the surface $z=z_\sigma$. Fitting the data to the expansion \eqref{eq:zsigmaHitExpansion}, we obtain $g_0$, $z_\sigma$, $f_{3/2}$ and $\psi_0$. With these data we then use the expansion \eqref{eq:zsigmaBounceExpansion} as our initial seed at $\tilde{z}=z_\sigma-10^{-20}$ and evolve towards $\tilde{z}=0$ stopping at a value where the singularity could be properly identified. For the plots shown here we found that stopping at $\tilde{z}=10^{-150}$ was enough to produce clear results.

\subsection{Extracting Kasner coefficients}\label{app: Numerical Kasners}
With the ansatz \eqref{eq:Ansatz}, it follows that if Kasner behavior is present at $z\rightarrow\infty$, we can compute the Kasner coefficients
\begin{equation}\label{eq:KasnerCoefsCompute}
p_t=\mathcal{N}\left(\frac{zg' }{2g}-1\right) \,,\qquad p_x=-\mathcal{N} \,,\qquad p_\psi=-z\,\mathcal{N}\partial_z\psi\,,
\end{equation}
where $\mathcal{N}=\left(zg'/(2g)-3\right)^{-1}$ and the above expressions have to be evaluated at $z\rightarrow\infty$. In practice, we use the above expression as a diagnostic of Kasner behavior: if these expressions stabilize for sufficiently large $z$ and satisfy the Kasner relations, we claim to find Kasner behavior.

\bibliographystyle{JHEP}
\bibliography{biblio}
\end{document}